\documentclass[fleqn,usenatbib]{mnras}

\usepackage[normalem]{ulem}
\usepackage[T1]{fontenc}
\usepackage{ae,aecompl}

\usepackage{graphicx}	
\usepackage{amssymb}	
\usepackage{hyperref}
\hypersetup{
  colorlinks   = true, 
  urlcolor     = blue, 
  linkcolor    = blue, 
  citecolor   = blue 
}
\usepackage{color}

\newcommand{\kms}{\,km\,s$^{-1}$}	
\newcommand{\caps}[1]{{\scshape{#1}}}

\title[The Intermediate Polar EX Hydrae]{A Radial Velocity Study of the Intermediate Polar EX~Hydrae}

\author[Echevarr\'{\i}a et al.]{J. Echevarr\'{\i}a$^{1}$\thanks{e-mail: 
	jer@astro.unam.mx; abdiel@ciencias.unam.mx ; rmm@astrosen.unam.mx; j.v.hernandez@soton.ac.uk }, 
	A. Ram\'{\i}rez-Torres$^{1\star}$, R. Michel$^{2\star}$ and J. V. Hern\'andez Santisteban$^{3}$\\
$^{1}$Instituto de Astronom\'ia, Universidad Nacional Aut\'onoma de M\'exico, 
Apartado Postal 70-264, \\Ciudad Universitaria, M\'exico D.F., C.P. 04510, M\'exico.\\
$^{2}$Instituto de Astronom\'ia, Universidad Nacional Aut\'onoma de M\'exico, 
Apartado Postal 877, \\Ensenada, Baja California, C.P. 22830 M\'exico.\\
$^{3}$Department of Physics \& Astronomy, University of Southampton, Southampton SO17 1BJ, UK
}

\begin{document}

\date{Accepted XXX. Received YYY; in original form ZZZ}

\pubyear{2016}


\pagerange{\pageref{firstpage}--\pageref{lastpage}}
\maketitle

\begin{abstract}
A study on the intermediate polar EX~Hya is presented, based on simultaneous photometry and high dispersion spectroscopic observations, during four consecutive nights. The strong photometric modulation related to with the 67-min spin period of the primary star is clearly present, as well as the narrow eclipses associated to the orbital modulation. Since our eclipse timings have been obtained almost 91,000 cycles since the last reported observations, we present new linear ephemeris, although we cannot rule out a sinusoidal variation suggested by previous authors. The system shows double-peaked H$\alpha$, H$\beta$ and \ion{He}{i} $\lambda$5876 \AA\@ emission lines, with almost no other lines present. As H$\alpha$ is the only line with enough S/N ratio in our observations, we have concentrated our efforts  in its study, in order to obtain a reliable radial velocity semi--amplitude. From the profile of this line, we find  two important components; one with a steep rise and velocities  not larger than $\sim$1000\kms and another broader component extending up to $\sim$2000\kms, which we interpret as coming mainly from the inner disc. A strong and variable hotspot is found and a stream-like structure is seen at times.  We show that the best solution correspond to $K_1 = 58 \pm 5$\kms\@ from H$\alpha$, from the two emission components, which are both in phase with the orbital modulation. We remark on a peculiar effect in the radial velocity curve around phase zero, which could be interpreted as a Rositter-MacLaughlin-like effect, which has been taken into account before deriving $K_1$. This value is compatible with the values found in high-resolution both in the ultraviolet and X-ray. Using the published inclination angle of $i =78 \pm 1\degr$ and semi-amplitude $K_2 = 432 \pm 5$\kms\@, we find: $M_{1} = 0.78 \pm 0.03$ M$_{\sun}$, $ M_{2} = 0.10 \pm 0.02$ M$_{\sun}$ and $a = 0.67 \pm 0.01$~R$_{\sun}$.  Doppler Tomography has been applied, to construct  six Doppler tomograms for single orbital cycles spanning the four days of observations to support our conclusions. Our results indicate that EX~Hya has a well formed disc and that the magnetosphere should extend only to about $3.75\,R_{\rm{WD}}$.

\end{abstract}

\begin{keywords}
stars: novae, cataclysmic variables  -- techniques: radial velocities -- eclipses -- ephemerides
\end{keywords}



\section{Introduction}

The short period intermediate polar (IP) EX Hydrae, belongs to the subclass of
cataclysmic variables in which the strength of the magnetic field of the white 
dwarf (WD) is not strong enough to achieve synchronization between the 
rotation of the primary and the orbital period of the binary \citep[see][and references therein]{war95}.

The WD accretes from a surrounding disc or ring into the magnetic poles, producing a variety of observable phenomena \citep[e.g.][]{mhl07a,beu08}. It was first identified by \citet{kra62} as an eclipsing system with an orbital period $P_{\rm{orb}} \sim 98 $ min, a disc inclination $i=78 \pm1\degr$ and a second prominent period $P_{\rm{spin}} \sim 67 $ min due to the rotation of the WD \citep{vog80,kru81}. The photometric visual light curve in EX~Hya is mostly dominated by the spin period which produces a sine-like modulation with an amplitude between $0.4 - 0.9$\,mag. There are also occasional enhanced maxima and persistent narrow eclipses clearly associated with the orbital period cycle \citep{vog80}. Earlier determinations of the radial velocity semi--amplitude of the primary, $K_1$, have been obtained in the optical by several authors \citep{bre80,cow81,gil82,hel87}. They estimated the masses of the binary from their derived $K_1$ values and the mass of the secondary from different mass-radius calibrations \citep{war76,rob76,pat84}. A first determination of $K_1$ from the FUV was obtained by \citet{mau99}. High resolution results have been obtained from the EUV and X-rays \citep{bel03,hoo04}. A first order estimation of the radial velocity semi-amplitude of the secondary, $K_2$, was derived by \citet{van03} using a cross-correlation technique and later by \citet{beu08} from \ion{Na}{i} and \ion{Ca}{ii} absorption lines.  Mass determinations for the primary star have also been obtained by \citet{fuj97} and \citet{cro98} from the X-ray spectra, as well from indirect methods by \citet{beu03}. A wide variety of mass values for the binary components has been obtained from these publications.

Its large $P_{\rm{spin}} / P_{\rm{orb}}$ ratio of $\sim2/3$ places this system out of the usual spin equilibrium rotation value of  $\approx 0.1$ \citep{kal91,kin99}. If this is the case, the magnetosphere must fill the Roche lobe of the primary star and there should be, at most, an accretion ring or no disc at all \citep[e.g.][]{kin99,mhl07a}. However, as pointed out by \citet{hel14}, there are several reasons to suggest that EX~Hya has a small magnetosphere far from equilibrium, among them are the eclipse timings of the partial X-ray eclipse which suggest a magnetospheric radius of only $\sim$ 4 times the white-dwarf radius; the lack of polarisation; the long-term secular decrease and some of the radial velocity studies, which show the presence of an emission component that is modulated with the orbital period. These controversies extend to the size of the inner disc.

The previous $K_1$ results have a large range of values. This may be due to the variety of methods used to determine this value, or due to the use of a combination of emission lines, or obtained with very different spectral resolutions. Since we have obtained high-resolution spectroscopy in the red and done simultaneous visual photometry, we believe that our data could help to unravel the reason of these different results. In particular, our analysis of H$\alpha$, includes a main disc as well as a broader accretion component. 

\section{Observations and Reduction}
\label{sec:observations}
 
\subsection{Photometry}

CCD photometry with the Johnson~V filter was obtained in 2008 January 10--13, using the 1.5-m telescope with the Marconi CCD for the first two nights and the 0.84-m for the rest of the nights with the Thomson detector (see Table~\ref{tab:LogPho}). All CCD images were processed using the IRAF\footnote{IRAF is distributed by the National Optical Observatories, operated by the Association of Universities for Research in Astronomy, Inc., under cooperative agreement with the National Science Foundation.} package. Images were bias subtracted and flat field corrected. A nearby comparison star with $m_V=11.39$~mag was used to determine the V magnitude of EX~Hya.

\begin{table}
\caption{Log of Photometric Observations}
\label{tab:LogPho}
\begin{center}
\begin{tabular}{cccccc}
\hline
      Date   &   &       &   Exp. & HJD (start)   &           HJD (end) \\ [-1ex]
      (UT 2008) & \raisebox{1.5ex}{Tel.}&\raisebox{1.5ex}{Filter} & (s) & (2454400+) &(2454400+)\\
\hline
	January 10 &  1.5-m    &    V  &     10   &   75.963844   &       76.010098\\
	January 11 & 1.5-m    &    V   &  10   &   77.009960      &    77.009752\\
	January 12 & 0.84-m    &   V  &     10   &   77.892149    &   78.060985\\
	January 13 & 0.84-m  &     V  &     10   &   78.899026     &  79.059909\\
\hline
\end{tabular}
\end{center}
\end{table}

\subsection{Spectroscopy}
\label{sec:spec}

Simultaneous spectroscopic observations of EX Hydrae was  obtained in 2008 January 10--13 with the Echelle spectrograph at the f/7.5 Cassegrain focus of the 2.1-m telescope of the Observatorio Astron\'omico Nacional at San Pedro M\'artir, B.C., M\'exico. The SITe3 1024$\times$1024 CCD was used to cover a spectral range  $\lambda\lambda4000 - 7100$~\AA  \, with a resolving power of R=12,000. An echellette grating of 300 ll/mm, was used and the exposure time was set to 240~s during the four nights. All CCD images were processed using the IRAF package. As the single spectra has a low signal to noise ratio, no flat fields were used and no sky subtraction was applied. This has no effect in the radial velocity analysis nor in the Doppler Tomography, but in the trail spectra there are faint traces of weak emission lines from the sky (see Section~\ref{sec:dopmap}).

\begin{table}
\caption{Log of Spectroscopic Observations}
\label{tab:LogSpec}
\begin{center}
\begin{tabular}{ccccc}
\hline
  Date &       HJD (start)	&  Time 	& No. of 	& Exp.	\\
  (2008 UT) &       (2454400+)	&  (hr) & spectra &  (s)\\
\hline
   January 10   & 75.962341  & 1.58	& 20	& 240\\
   January 11   & 77.027255  & 1.26	& 15	& 240\\
   January 12   & 77.950838  & 2.65	& 32	& 240\\
   January 13   & 78.942922  & 2.64	& 32	& 240\\
\hline
\end{tabular}
\end{center}
\end{table}

\section{Photometric Analysis}
\label{sec:photo}

\subsection{Spin modulation and orbital eclipses}
\label{sec:mod}

\begin{table}
\caption{Eclipse and spin timings}
\label{tab:timings}
\begin{center}
\setlength{\tabcolsep}{3\tabcolsep} 
\begin{tabular}{ccc}
\hline
   Cycle	   &  HJD - 2454400  &   O-C \\
\hline
 \hspace{-3ex}Orbital & & \\
 245861	 &    75.981947	  	& 	 -0.000360	\\		
 245877	 &    77.073114	  	& 	 -0.000934	\\		
 245889	 &    77.892149	  	& 	 -0.000705	\\		
 245890	 &    77.960593	  	& 	 -0.000495	\\		
 245891	 &    78.028783	  	& 	 -0.000539	\\		
 245904	 &    78.915660	  	& 	 -0.000702	\\		
 245905	 &    78.983698	  	& 	 -0.000898	\\			
 245906	 &    79.052409	  	& 	 -0.000421	\\[1ex]
 \hspace{-5ex} Spin & & \\	
360418	 &	75.998373	 &	0.000530	\\ 
360440	 &	77.019359	 &	-0.002497\\
360441	 &	77.063692	 &	-0.004710\\
360459	 &	77.907324	 &	0.001094\\
360460	 &	77.951032	 &	-0.001744\\
360461	 &	78.000574	 &	0.001252\\
360462	 &	78.047882	 &	0.002013\\
360481	 &	78.93396	    	 &	0.003717\\
360482	&	78.974901	&	-0.001888 \\
360483	&	79.023783	&	0.000448  \\
\hline
\end{tabular}
\end{center}
\end{table}

\begin{figure*}
 \begin{center}
\includegraphics[trim=10 20 0 0,width=2.2\columnwidth]{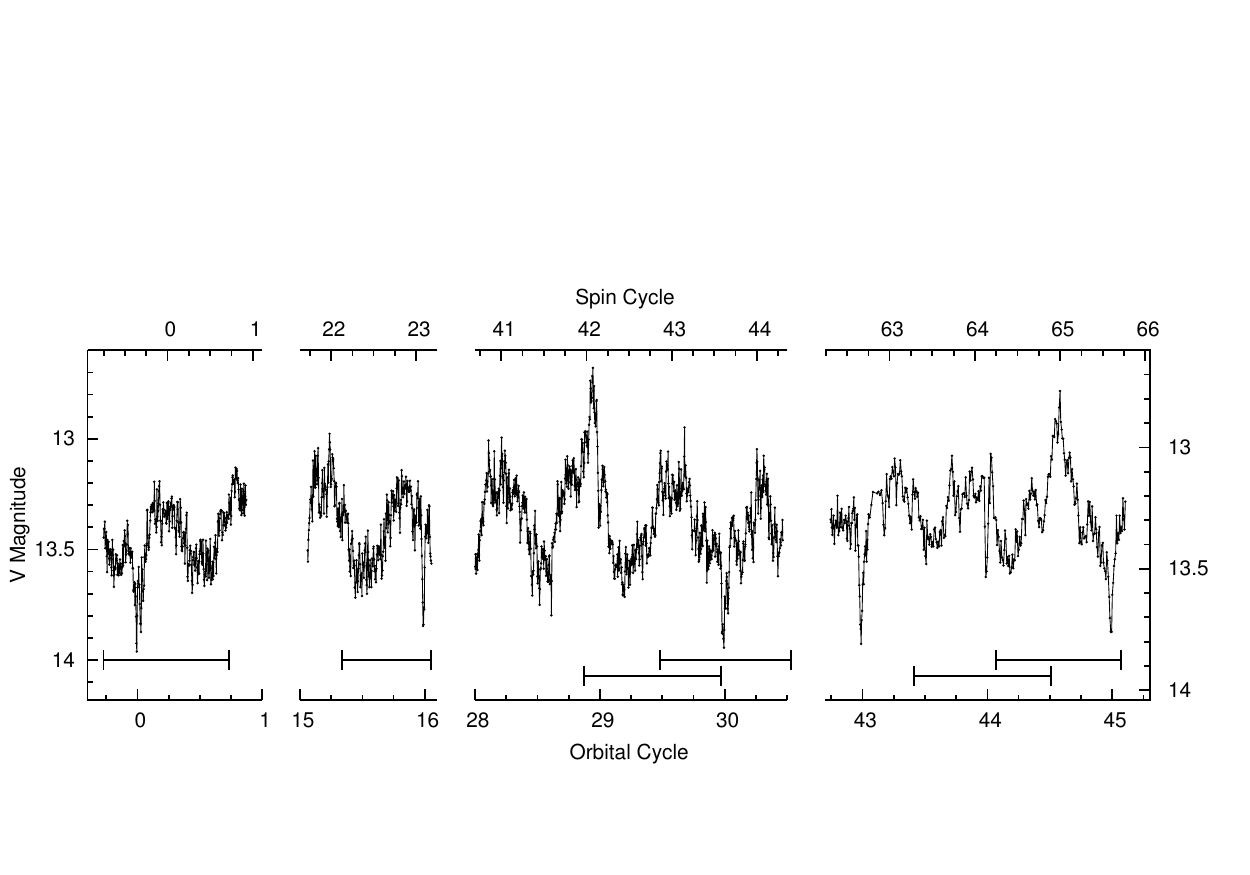}
\caption{V~Johnson photometry of EX~Hya during four consecutive nights. The light curves show narrow partial eclipses, not always at  the minimum of the light curve, since this is strongly modulated with the spin cycle as evident in the lower and upper axes. The horizontal bars indicate the simultaneous spectral coverage corresponding to a single tomogram (see Section~\ref{sec:dopmap}).}
\label{fig:vphot}
\end{center}
\end{figure*}

With our photometric data, we were able to measure eight eclipses and ten spin maxima.Their timings are shown in Table~\ref{tab:timings}. The O-C values have been obtained with our new calculated orbital ephemeris (see Section~\ref{sec:ephem}) and with the cubic spin ephemeris recently calculated by \citet{mau09}. The light curves of the observed object  are shown in Figure~\ref{fig:vphot}. EX~Hya shows a mean V~magnitude of $m_V = 13.3$\,mag and narrow partial eclipses of $0.4$\,mag, associated with the orbital cycle. The orbital period is shown in the bottom axis and is set to zero for the first eclipse, based on our ephemeris. As we can see, the subsequent eclipses coincide with integer values of the orbital period. The light curves are dominated mostly by the spin period modulation, as clearly shown in the top axis, where we have marked the spin cycles in a similar manner as in the orbital period. The modulation shows amplitudes between 0.5 to 0.8\,mag, with occasional enhanced maxima (e.g. spin cycles 42 and 65). The eclipses are clearly seen, independently of the strength of the spin modulation. This is particularly true in orbital cycles 29 and 44. The total difference between the enhanced maxima (spin cycle 42) and the bottom of the last eclipse (orbital cycle 45) is 1.1\,mag. We find a very similar behaviour to that found by \citet{vog80}. Comparing with the AAVSO\footnote{\url{https://www.aavso.org}} light curve, EX~Hya  was observed at a low state, long before the outburst detected in May 2010.

\begin{figure*}
\begin{center}
	\includegraphics[trim=0 0 0 0,angle=270,width=2.2\columnwidth]{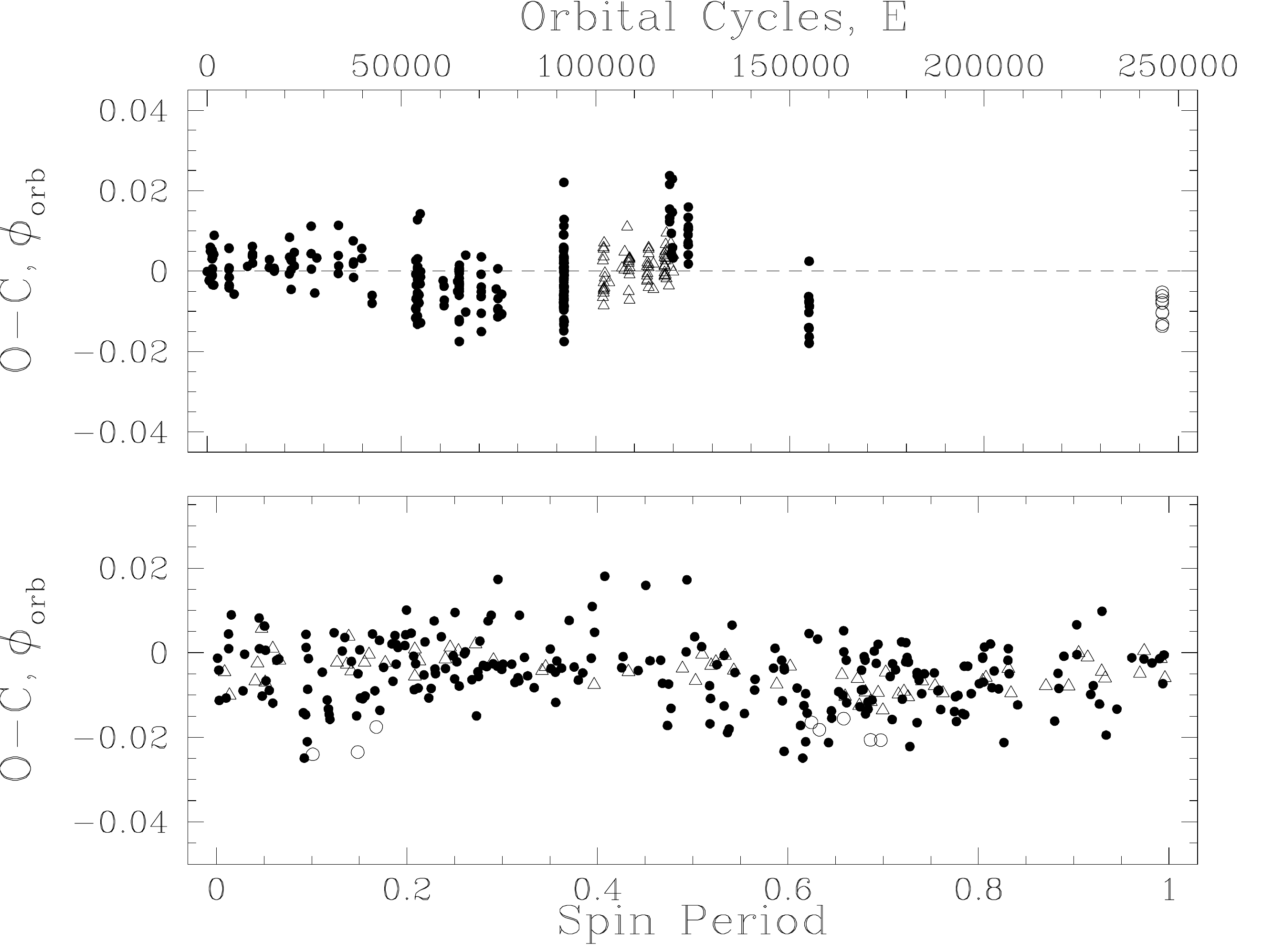}
	\caption{\emph{Top}: O-C residuals in orbital phase versus cycle number for 342 eclipse minima using the new ephemeris showing an initial dispersion around zero. \emph{Bottom}: All the eclipses from the top panel are folded by the spin period, using the ephemeris by \citet{mau09}. The open triangles are the observations by \citet{jab85}, the open circles are the minima obtained in this paper. We note that a sinusoidal behaviour with the spin do not prevail in the long term (see text).}
\label{fig:omc}
\end{center}
\end{figure*}

\citet{hea85} noted a possible sinusoidal variation of the orbital period, which also appears to be present in the results by \citet{baf88} and \citet{hel92}.  

\subsection{New ephemeris}
\label{sec:ephem}

Since these last observations, no further eclipse timings have been published until now. Close to 91,000 orbital cycles have since elapsed. For this reason, we have  decided to review the available data of the published eclipses. We have compiled 342 eclipses since the first observations in 1962 \citep[see,][]{mum67}, all of which are now available in digital form\footnote{Available at the CDS via anonymous ftp to \url{cdsarc.u-strasbg.fr} (130.79.128.5) or via \url{http://cdsarc.u-strasbg.fr/viz-bin/qcat?J/MNRAS}}. We have calculated new linear orbital ephemeris with the following results:

\begin{equation}
HJD_{\rm{eclipse}} = 2,437,699.94131(11) + 0.068233843(1) E.
\end{equation}

The O-C residuals versus orbital cycles using these ephemeris are plotted in Figure~\ref{fig:omc}. Although the observations show a positive trend between cycles 120,000 - 130,000 and a possible sinusoidal variation may be present, our data contributes little to support this oscillation. Another possibility is that the sample of eclipse observations could produce an unwarranted O-C bias. The observations by \citet{sea83}, which cover 14 consecutive days between cycles 91,729 to 91,927 show that, with enough eclipse timings, the O-C has a large scatter centred around zero. In other runs (including ours, which cover only four nights and have a small negative O-C value) the bias could arise from the small number of eclipse observations. The origin of the scatter in the O-C could be the result of a variable eclipse timing, which could depend on the location of a hotspot. 
 
In contrast, \citet{sie89} present the analysis of 25 eclipses (observed in March 1983) in addition to the observations of \citet{jab85}, they found a sinusoidal behaviour locked to the spin period. They explained this modulation as an optical eclipse wandering back and forth by $\pm$20 s as a function of the spin phase, from which they infer a source close to the white dwarf and at a distance of about 2~$R_{\rm{WD}}$. However, this modulation does not prevail in the long term, as seen in the nearly fifty years of eclipse observations shown in the bottom panel of Figure~\ref{fig:omc}. Here, we plot the observations by \citet{jab85} (open triangles) and ours for reference (open circles). Together with the rest of the points, we used the orbital ephemeris by \citet{mum67} to compute the O-C values, for direct comparison with Figure 3 of \citet{sie89} and the cubic spin ephemeris by \citet{mau09}. We were unable to include the eclipses from \citet{sie89}, since they do not publish their eclipse timings. 

\section{Spectroscopic Analysis}
\label{sec:spec}

\subsection{A comparison of EX~Hya with previous spectroscopic studies in the optical}
\label{sec:comp}

\begin{table*}
\caption{Reported semi-amplitude values in radial velocities analysis and derived masses}
\label{tab:k1lit}
\begin{center}
\setlength{\tabcolsep}{2\tabcolsep} 
\begin{tabular}{cccccccc}
\hline
      $K_1$  &      $K_2$ & spectral line & resolving power & $M_1$ & $M_2$ &References\\
     \kms &\kms  & & R &  M$_{\sun}$ & M$_{\sun}$ \\
     \hline
	 $68 \pm 9$      &   --     & H$\beta$, H$\gamma$, H$\delta$	& 124\,\AA /mm$^{a}$ & 1.4 & 0.19 & \cite{bre80}  \\ 
	 $90 \pm 28$    & --	      &H$\beta+$H$\gamma+$H$_8$ 	& 47\,\AA /mm$^{a}$  & 0.7 & 0.16 & \cite{cow81}   \\
	 $58 \pm 9$      &  --	& H$\gamma+$H$\delta$ 	& 2100 & 1.4 & 0.17 & \cite{gil82}     \\
	 $69 \pm 9$      &   -- 	& H$\beta+$H$\gamma$ 	& 4600  & 0.78 & 0.13 & \cite{hel87}   \\
	$85 \pm 9$       &    --	& \ion{O}{vi} (FUV)		& 3000 & -- & -- & \cite{mau99} \\
	$59.6 \pm 2.6$ &  --	& \ion{N}{v}, \ion{O}{v} (FUV) & 45,800 & 1.33 & 0.15 & \cite{bel03}   \\
	$58.2 \pm 3.7$	&  --      & CLP$^b$  (X-ray) 		& 20,000 & 0.49 & 0.08 & \cite{hoo04} \\
	     --                 & $360 \pm 35$ & \ion{Na}{i}  & 120 & 0.47 & 0.10 & \cite{van03} \\
	     --                 & $432 \pm 5$ & \ion{Na}{i}, \ion{Ca}{ii}  & 47,000  & 0.79 & 0.11 & \cite{beu08} \\
	$58 \pm 5$      &    --                & H$\alpha$  & 12,000  & 0.78 & 0.10 & Present Paper \\

\hline
\end{tabular}
\end{center}
\raggedright
$^a$ Observations with photographic plates. \\
$^b$ Composite Line Profile for Fe, S, Si, Mg and Ne lines in the X-ray regime.
\end{table*}

As shown in Table~\ref{tab:k1lit}, there are several radial velocity studies and mass determinations for this binary. Most of them have been focused on obtaining the semi-amplitude of the primary star, $K_1$. The measurements have been performed in a wide wavelength range: four in the optical, two in the UV and one in the X-ray region. In contrast, only two studies of the secondary  have been published (due to the difficulty to observe the secondary in short orbital period systems). The four previous optical studies, which concern the purpose of this paper, are based on the blue part of the optical spectrum, which show strong hydrogen Balmer lines up to H$\beta$. Their results are based mainly on different combination of lines as shown in the Table \ref{tab:k1lit}. To our knowledge, no published work has been done in the red part of the optical spectrum.

 \begin{figure}
  \begin{center}
    \includegraphics[angle=0.0,trim=0cm 0cm 0cm 0cm,clip,width=1.0\columnwidth]{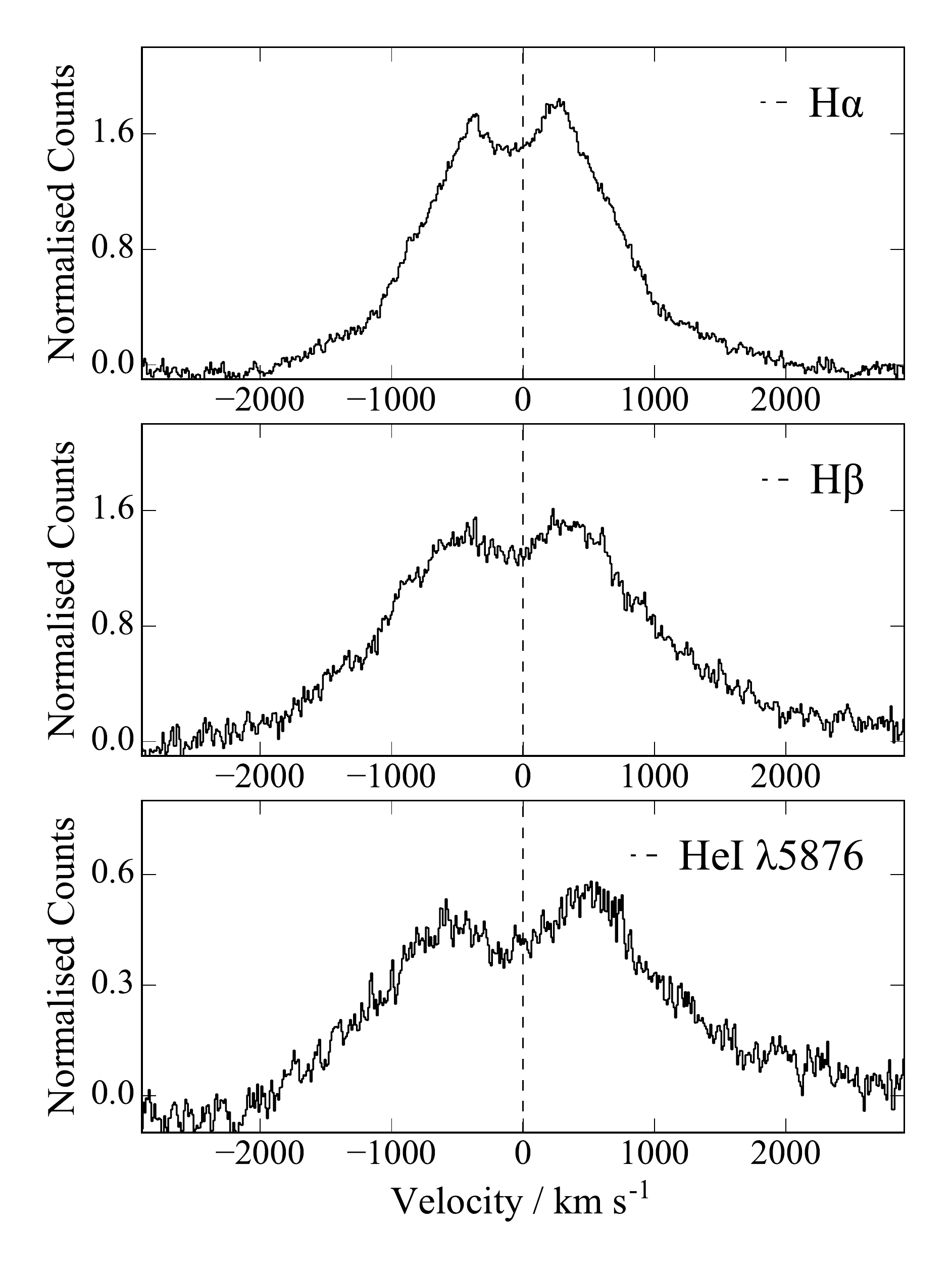}
    \caption{Co-added spectra of H$\alpha$, H$\beta$ and HeI$\lambda$ 5876 emission lines. The spectra have normalized counts at a continuum level fourty angstroms away from the nominal line centre.}
    \label{fig:em-lines}
  \end{center}
\end{figure}

Our Echelle observations cover the region $\lambda\lambda$4000--7100\,\AA.  Our spectra do not show prominent higher-order Balmer lines. In the red region, we observe double-peaked H$\alpha$, H$\beta$ and \ion{He}{i} $\lambda$5876 lines (see Figure~\ref{fig:em-lines}). However, we do not detect the high-excitation line \ion{He}{ii} $\lambda$4686. 

This might suggest that we observed EX~Hya at a lower excited state, but we do not rule out that our spectra are affected by a large atmospheric extinction since the spectra were observed through relatively high air masses (with values in the range 2.0 to 2.8), and the lack of higher Balmer lines might be the result of high extinction in the blue part of the spectrum. 

Due to the short exposure times used to cover enough spectra over an orbital period, only the strongest H$\alpha$ line has a strong S/N ratio suitable for a radial velocity study (our attempts on $H\beta$ give very poor results). \ion{He}{i} is even weaker and is seen only in the co-added spectrum (see Figure~\ref{fig:em-lines}). The H$\alpha$ trail spectra presents strong S-wave modulation (see Section~\ref{sec:dopmap}) as well as a double-peaked emission presumably coming from the disc and a low intensity broad component. The latter has been interpreted by \cite{hel87} as emission originated from the accretion curtain. We note a change in slope of the wings of the line at $\pm$1000 \kms. We will refer to the emission within this velocity range as the main component. In the next subsections, we will derive the radial velocity semi-amplitude for the wings of the main component using a standard double Gaussian technique and then, the radial velocity of the broad component only using a masking technique. We should state here that in all our work done in the next Sections, we have used the 99 individually spectra available to us. We believe that this will give us a better temporal and/or phase dependent changes. We will discuss this approach, instead of using a binning process in Section~\ref{sec:discus}.

\subsection{Radial Velocity - Double Gaussian}
\label{sec:ws-code}

To obtain the orbital parameters of the white dwarf from H$\alpha$, we assume that the double-peaked emission line comes from a symmetric accretion disc and derive the individual radial velocities using the standard double-Gaussian technique and its diagnostic diagram as described by \citet{sha86}. We have used the \emph{convrv} routine (kindly shared with us by Thorstensen, private communication), within the IRAF \emph{rvsao} package, to compute radial velocities. In order to give a stronger weight to the wings of the double-peaked emission line, we have used the {\it gau2} algorithm which fits the line profile with two Gaussians of the same fixed FWHM. A preliminary search was made using 6\,\AA-width Gaussians. To determine the optimal separation  between the two gaussians, we fitted, for a wide selection of separation values {\bf \emph{a}}, the set of radial velocities  to a circular orbit:

\begin{equation}
\label{eq:velrad}
V(\phi) = \gamma + K_{\rm{em}}~\sin~[2\pi (t - HJD_{\odot})/ P_{orb}] ,
\end{equation}

\noindent where $V(\phi)$ are the observed radial velocities, $\gamma$ is the systemic velocity, $K_{\rm{em}}$, are the corresponding semi amplitudes derived from the radial velocity curve, $HJD_{\odot}$ is the heliocentric Julian Date at the inferior conjunction of the secondary star, and $P_{orb}$ is the orbital period of the binary. 

\begin{figure}
  \begin{center}
    \includegraphics[width=\columnwidth]{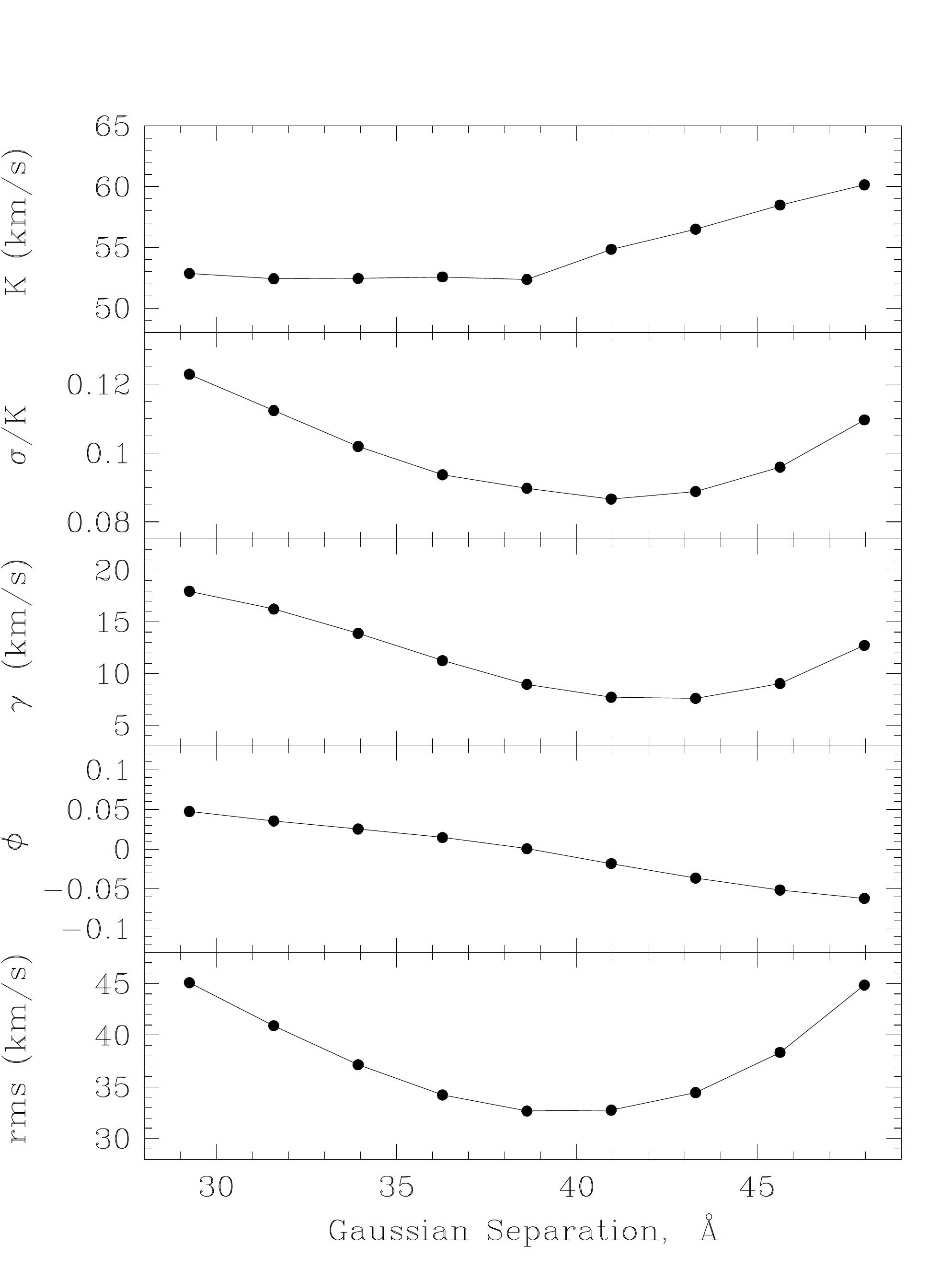}
    \caption{Diagnostic Diagram using the double Gaussian method for the estimation 
    of the orbital parameters varying the separation between the Gaussians in steps of 1\,\AA\@ with a fixed width of 6\,\AA\@
    (see text for a full description).}
    \label{fig:diagnostic1}
  \end{center}
\end{figure}

The result of the diagnostic diagram is shown in Fig.~\ref{fig:diagnostic1}. The best orbital parameter solution was selected by defining the minimum  in $\sigma/ K_1$ value used as the best-fit parameter. We found an optimal Gaussian separation of $a = 41$\,\AA which corresponds to a semi-amplitude of $K_1= 55 \pm 5$\kms. The overall results are satisfactory and we find a $K_1$ compatible with the UV and X-ray results. However, we find inconsistencies in the diagnostic diagram that should be addressed. As pointed out by \citet{sha86} in their study of SW~UMA, an intermediate polar with an orbital period (81.8 min) similar to EX~Hya the phase-dependent asymmetries should be readily identified in the diagnostic diagram. $K_1$ should approach a stable value as the Gaussians separation $a$ becomes sufficiently large, and $\sigma/K_1$ should have a significant increase at large values of $a$ indicating that the velocity measurements become to be dominated by the noise at the continuum. Furthermore, the phase shift should flatten out, also at large values of $a$, if the asymmetry is confined to low velocities. In our case, these indicators do not behave in such a manner,  $\sigma/K_1$ has only a moderate increase, $K_1$ does not show a stable value at any separation $a$, nor does the phase shift becomes stable at large values of $a$. We are concerned also that the radial velocity values show a strong deviation from a sinusoidal behaviour (not shown in the paper), which could indicate that we have not properly avoided all phase-dependent asymmetries. In the diagnostic diagram we have included the rms behaviour with $a$. Although this parameter is a  general measure  of the overall four-parameter orbital solution, we find it useful to check on the overall results. In fact, the rms value has a minimum, consistent with our $\sigma/K_1$ and $K_1$ results. The slow upward change in $\sigma/K_1$ might be due to the fact that the broad wings are still contributing to an orbital solution, but because they have a less steep slope than the stronger central wings, the double Gaussian method might not yield the best solution. With respect to the behaviour of the diagnostic diagram indicators we must point out that this is not a matter of selecting larger values of $a$. The maximum separation of 48\,\AA  ~is due to the fact that at larger values the \emph{convrv}  routine breaks down.  Our choice of using 6\,\AA ~Gaussian widths might also be contributing to the above-mentioned inconsistencies; although the choice of widths is usually not a problem, provided that these widths are much narrower than the separation of the Gaussians.

\begin{figure}
  \begin{center}
    \includegraphics[width=1.0\columnwidth]{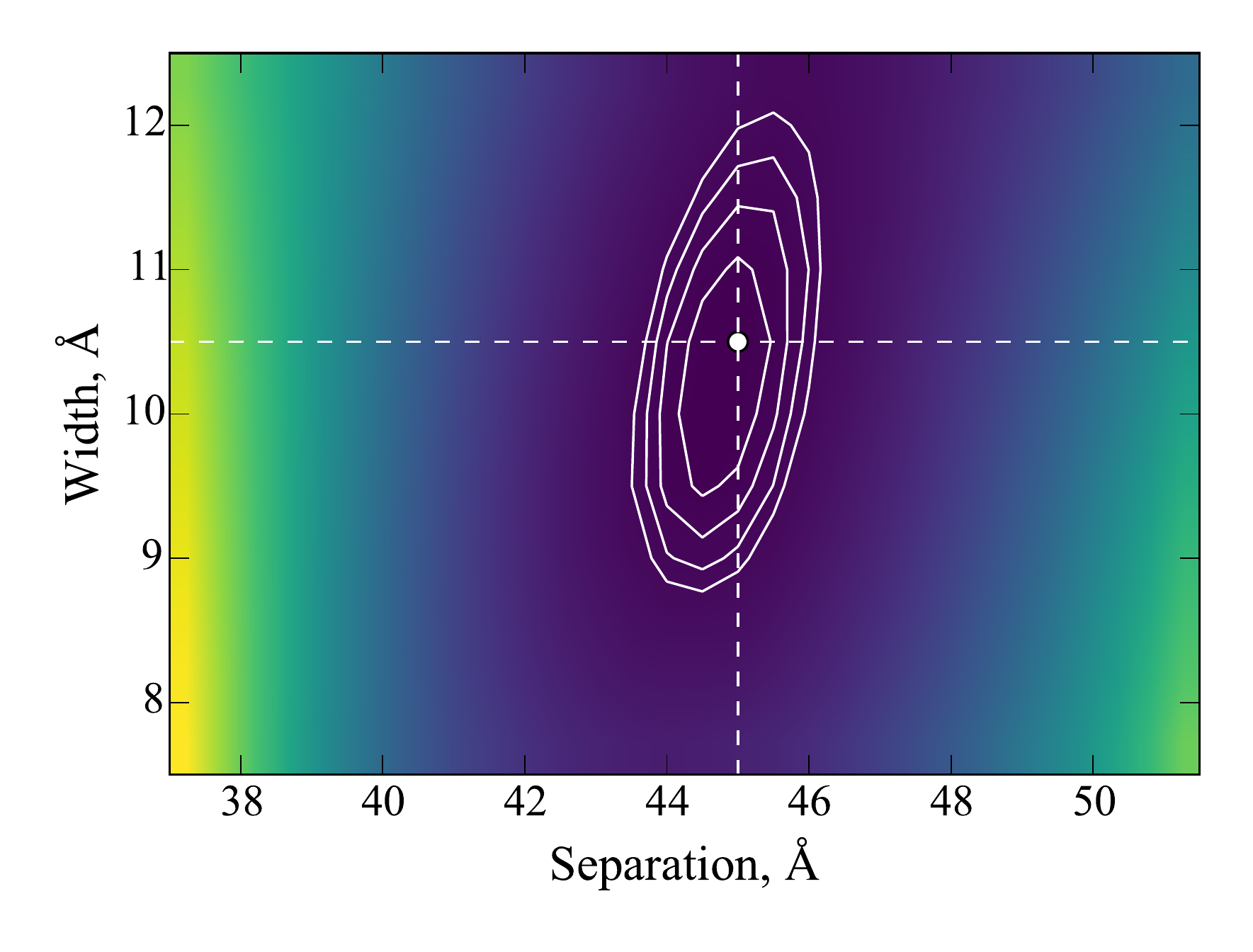}
\caption{The two dimensional map used to select the best double Gaussian fit with a  separation of 45\,\AA\, and a width of 10.6 \,\AA. Contours represent $\Delta(K_1/\sigma)=0.001$ steps. Colours show $K/\sigma$ from 14 (deep blue) to  7 (light yellow).}
    \label{fig:2-d}
  \end{center}
\end{figure}

For these reasons, we decided to find the optimal Gaussian width and separation interactively. One of us (JVHS) developed a \caps{python} code to produce an interactive width-separation programme. The code uses a grid in which the width and separation can vary iteratively for a wide selection of values, and fits, in every trial, the four parameters  set for the circular case, given in Eq. \ref{eq:velrad}. In our case we set the interactive grid from 5.5 to 13\,\AA ~in 0.1\,\AA ~steps and the separation from 37 to 51.5\,\AA ~in 0.1\,\AA  ~steps and found an optimal width for the Gaussians at 10.6\,\AA~and separation~$a$ of 45\,\AA.  A two-dimensional map is shown in Figure~\ref{fig:2-d} to illustrate the selection approach.  The contours represent $\Delta(K_1/\sigma)=0.001$ steps.

\begin{figure}
  \begin{center}
    \includegraphics[width=1.0\columnwidth]{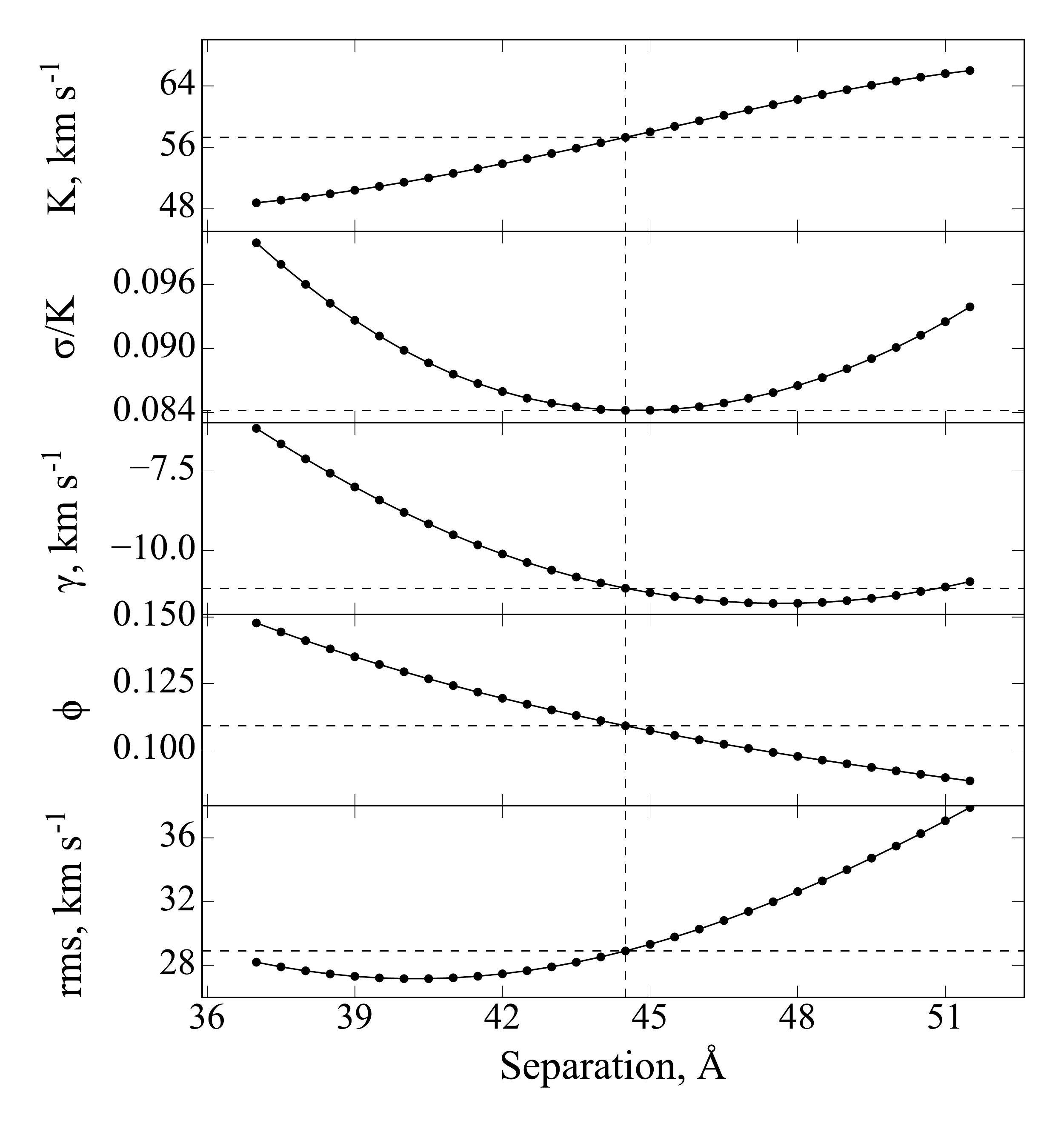}
    \caption{Diagnostic Diagram using the double Gaussian method  with the interactive width-separation program. The optimal  values for the width and separation are discussed in the text. Note that we are presenting the results of Case B  (see text for a full discussion on the diagram and best orbital solutions).}
    \label{fig:diagnostic3}   
   \end{center}
\end{figure}

In Figure~\ref{fig:diagnostic3}, we present a diagnostic diagram to determine the best orbital solution by finding the minimum in the control parameter $\sigma / K_1$. The optimal value is shown as a dotted vertical line. Although now  $\sigma/K_1$ appears to increase rapidly for high $a$ values, the fact is that this is only a visual effect in the diagram, the increase is still small as in the previous diagram. Furthermore, $K_1$ and the phase shift still shows no stable values at any separation $a$ and the rms residuals do not follow the control parameter $\sigma/K_1$, reaching its minimum value for lower separations and increases constantly even before the $\sigma / K_1$ minimum is reached. The new method still does not avoid the possible fact that the broad wings may be affecting the best orbital solution, i.e. the double Gaussian method might not yield the best solution, even if we have solved now the limitation of the maximum separation of 48\,\AA \, imposed by the \emph{convrv}  routine. This will be addressed again by examining the broad component alone in Section~\ref{sec:radvel2}.

\begin{figure}
 \begin{center}
\includegraphics[trim=0 0 0 0,clip,width=1.05\columnwidth]{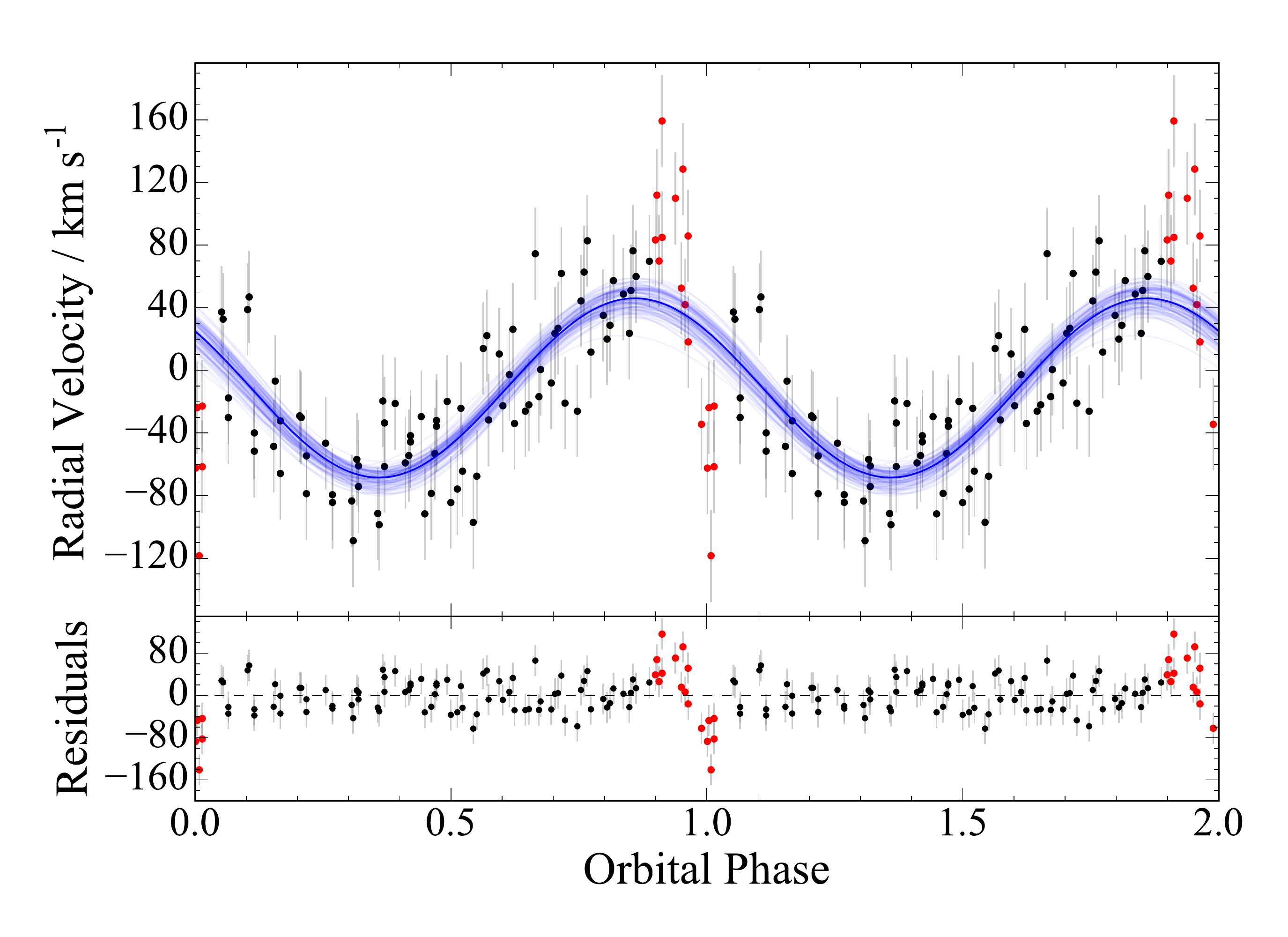}
\caption{Radial velocity curve for H$\alpha$ emission line and the best solution from the interactive width-separation programmme (see text). We show the results for Case B, where the red points have been excluded from the fit (see Section~\ref{sec:rm-type} for details). Errors on the individual data points have been scaled so $\chi^2_{\nu}=1$. Random realizations via bootstrapping are shown (\textit{blue lines}) to reflect the scatter of the solution.}
\label{fig:velcur1}
\end{center}
\end{figure}

The best orbital solution is shown in Figure~\ref{fig:velcur1}. The orbital parameters errors were calculated by computing 1000 bootstrap copies of the radial velocity curve and repeating the fitting process. The bootstrap distributions of each parameter are well described by Gaussians, therefore we used the mean and standard deviation to determine the value and the Gaussian sigma. We repeated the same process after excluding those data points whose values deviate from the sinusoid (shown in red) and are discussed in Section~\ref{sec:rm-type}. We summarised the results in Table~\ref{tab:OrbParam2} where we call 2-Gaussian - Case A (all points) and Case B (excludes the red points), respectively. 
\begin{table*}
\caption{Orbital Parameters of EX~Hya obtained with the interactive double Gaussian grid and for a single Gaussian with a mask. Case A and B show the solutions with and without the RM-type points, respectively (see text for full discussion).}
\label{tab:OrbParam2}
\begin{center}
\setlength{\tabcolsep}{.32\tabcolsep} 
\begin{tabular}{l l c c c c c c c c}
\hline
      \multicolumn{2}{l}{Orbital } & \multicolumn{4}{c}{2-Gaussian} & \multicolumn{4}{c}{Core Mask} \\
      \multicolumn{2}{l}{Parameters} & \multicolumn{2}{c}{A}& \multicolumn{2}{c}{B} & \multicolumn{2}{c}{A}& \multicolumn{2}{c}{B}  \\
\hline
	$\gamma$  &\kms 	  &  &-13 $\pm$ 4 &  &-19 $\pm$ 4 &    &14 $\pm$ 4   &    &13 $\pm$ 4     \\
	$K_1$ & \kms          	&  &60 $\pm$ 5   &  &57 $\pm$ 5 & &62 $\pm$ 6 & &58 $\pm$ 5      \\
	HJD$_{\odot}$  & (+ 2454475 days)  &      &0.989 $\pm$ 0.01   &  &0.987 $\pm$ 0.001 &     &0.987 $\pm$ 0.001& &0.988 $\pm$ 0.001   \\
	rms    &      \kms 	   &    & 37&    & 36&  &  38&  &  30 \\
\hline
\end{tabular}
\end{center}
\end{table*}

\subsection{Radial Velocity of the broad component: Masking of the Core}
\label{sec:radvel2}

\begin{figure}
 \begin{center}
\includegraphics[trim=0 0 0 0,clip,width=1.05\columnwidth]{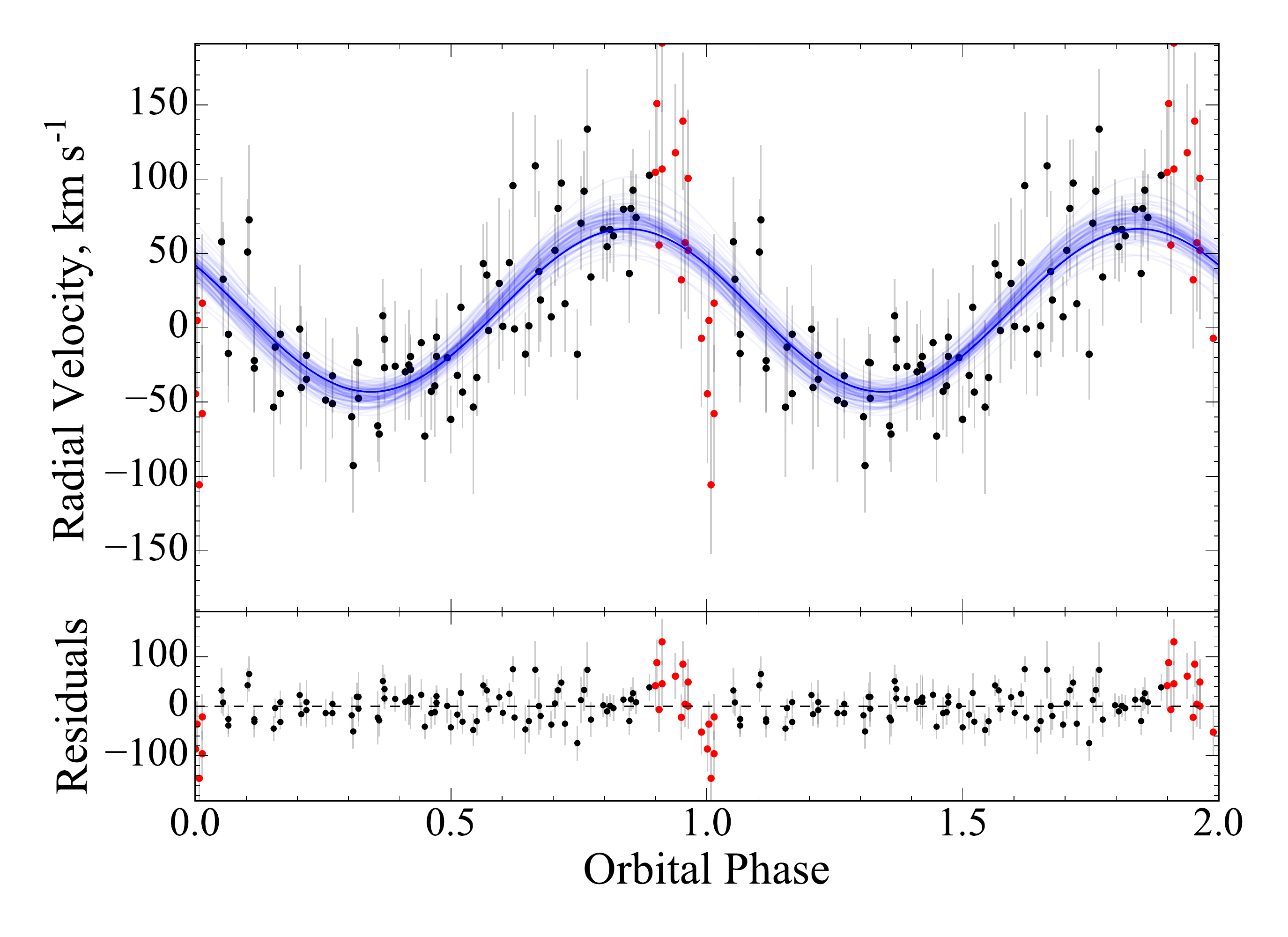}
\caption{Radial velocity curve for H$\alpha$ broad emission line component with the best solution by masking the core of the line.  We used the results from the core mask method (case B). Red points have been excluded from the fit (see Section~\ref{sec:rm-type} for details). Individual errors on the each data point have been obtained  now from the individual gaussian fits to each spectrum and scaled again so $\chi^2_{\nu}=1$. Random realizations via bootstrapping are shown (\textit{blue lines}) similar to those obtained in Figure~\ref{fig:velcur1}.}
\label{fig:velcur2}
\end{center}
\end{figure}

It is possible that the broad wings are still been contaminated by the core of the line. Therefore, we explored this possibility by performing a fit to the broad and high velocity wings only while masking the core of the line. For every spectra taken, we fitted a single Gaussian to obtain an initial estimate of the line center. Then, we applied a 34\,\AA\@ mask around this initial estimate and repeated the Gaussian fit. This allowed us to symmetrically fit the high velocity component ($1000 - 2000$\kms) of the line at each individual phase. We have calculated the individual errors for each radial velocity measurement by bootstrapping 1000 copies of the masked spectra and performing the Gaussian fit. In order to propagate errors, every set of bootstrapped radial velocities were used to calculate the errors on the orbital solution by fitting Eq. \ref{eq:velrad}. Each orbital parameter distribution is well described by a Gaussian and therefore we use the mean and standard deviation as the cited value and the Gaussian sigma. The best solution for the radial velocity is shown in Figure~\ref{fig:velcur2}.  The results are also shown in Table~\ref{tab:OrbParam2}, which we call Core Mask  Case A (all points) and Case B (without the asymmetric points discussed in the next Section), respectively, which are almost identical to those obtained with the double Gaussian method. Note that all times of inferior conjunction HJD$_{\odot}$ are about 0.1 in phase ahead of the eclipse timings derived in our ephemeris. This is the case in many disc CVs, where the hot spot lies ahead (e.g. \citet{war95}) and in our case is consistent with our tomographic results, i.e. the eclipses are a partial occultation of the hot spot. Note also that all four solutions give a comparable results within the errors. This is probably due to the fact that the red points are only a few, and they are fairly symmetric in high and low velocities. We select the result for the Mask Core Case B as it has the lowest rms value. However, using Case B for the Double Gaussian will not change our results. We will therefore use $K_1 = 58 \pm 5$\kms which is, within the errors, consistent with the UV and X-ray results.

\subsection{A Rossiter-McLaughlin-type effect}
\label{sec:rm-type}

The asymmetric behaviour in the radial velocities is clearly present between phases $\sim$0.8 and 0.9. (red points in Figures~\ref{fig:velcur1} and \ref{fig:velcur2}. The radial velocities show first a substantial excess followed by a sudden drop to negative velocities. The radial velocities then resume a normal sinusoidal behaviour (within the errors). This anomaly resemble the Rositter-McLaughlin (RM) effect, seen in many occulting binaries since its discovery in $\beta$ Lyrae by \citet{ros24} and Algol \citet{mac24}, and explained as an effect of rotation during the eclipse in the velocity of the eclipsed component. In our case, we will show in Section~\ref{sec:discus} that a similar effect can bee seen by en eclipse of a symmetrical component, rotating close to the white dwarf. It is important here to mention that these asymmetries are present in the data of the previous optical works \citep{bre80,cow81,gil82,hel87} and they have been interpreted as an occultation by the companion star of a mass-flow that is circulating around  the white dwarf by \citet{cow81} and \citet{hel87} (the latter even exclude their points around phase zero in the calculation of $K_1$). It is difficult to directly compare these asymmetries with ours, due to the different method and the variety of lines used by these authors to describe this effect. Since we have enough time resolution on a single line, we can see the asymmetry directly on our radial velocity curve, which we will refer hereinafter as an RM-type effect. The velocities showing this effect are shown in red in Figure~\ref{fig:velcur1} and \ref{fig:velcur2}. It is important to note that the RM-type points have a close symmetry regarding positive and negative values extending to roughly $\pm 150$ \kms. Further discussion on the location of the occulted source will be addressed in Section~\ref{sec:discus}.


\subsection{Basic system parameters} 
\label{sec:baspar}

Assuming that the radial velocity semi-amplitudes reflect accurately the motion of the binary components, then from our result, $K_{\rm{em}}= K_1 = 58 \pm 5$\kms (see Section~\ref{sec:discus} for our choice of K value), and adopting $K_{\rm{abs}}=K_2=432 \pm 5$\kms \citep{beu08} and  $P_{\rm{orb}}=0.068233846$ from our new ephemeris (see Section~\ref{sec:ephem}), we obtain:

\begin{equation}
q = {K_1 \over K_2} = {M_2 \over M_1} = {0.13 \pm 0.02},
\end{equation}

\begin{equation}
M_1 \sin^3 i = {P K_2 (K_1 + K_2)^2 \over 2 \upi G} = 0.73 \pm 0.03\, \rm{M}_{\sun},
\end{equation}

\begin{equation}
M_2 \sin^3 i = {P K_1 (K_1 + K_2)^2 \over 2 \upi G} = 0.10 \pm 0.01\, \rm{M}_{\sun},
\end{equation}

\noindent and

\begin{equation}
 a \sin i = {P (K_1 + K_2) \over 2 \upi} = 0.66 \pm 0.01\, \rm{R}_{\sun}.
\end{equation}

Since the orbital eclipses on EX~Hya are shallow and overpowered by the spin modulation, no eclipse modeling has been applied to determine the inclination angle. Therefore, its value has been determined by previous authors by assuming a mass-radius relation (e.g. \citet{bre80,hel87}), or by the assumption that the X-ray emission is point-like centered on the white dwarf \citep{hoo04,beu08}. However, most authors agree on a value of  $i =78\degr \pm 1$ \citep[e.g.][and references therein]{war95}. In any case, a small change in $i$ at this high inclination would exert little effect on the system parameters. Using this value for $i$ we obtain: $M_{1} = 0.78 \pm 0.03$ M$_{\sun}$, $ M_{2} = 0.10 \pm 0.02$ M$_{\sun}$ and $a = 0.67 \pm 0.01$ R$_{\sun}$.

Within the uncertainties, we find a compatible $K_1$ value as those obtained in the ultraviolet by \citet{bel03} and in the X-rays by \citet{hoo04} (see also Table~\ref{tab:k1lit}). This points towards the fact that there is indeed an accretion disc, whose inner parts reflect the motion of the white dwarf. Our results are compatible with those of \citet{beu08}, who derives an accurate semi-amplitude for the radial velocity of the secondary and uses the ultraviolet and X-ray results. Thus, this is the first time that we are able to derive a confident $K_1$ value from the optical. Although \citet{gil82} reports also a compatible value, he averages three very different semi-amplitude results obtained from a combined H$\delta$+H$\gamma$ fit using three different wing velocity regions on the same data. Furthermore, his results on H$\beta$ give $K_1 = 71.6 \pm 16.0$ even though its intensity is greater than the ones corresponding to H$\gamma$ and H$\delta$.

The masses of the primary star, derived by the authors shown in Table~\ref{tab:k1lit} have a large range in values, with values as low as 0.47 M$_{\sun}$ and up to almost the Chandrasekhar limit. This large range is probably due to the uncertainty and variety of methods used in deriving $K_1$. However, short orbital period systems, like EX~Hya, should have secondary masses close to those of main sequence stars \citep{war95,ech83,bea98,kni06}. Using a mass-radius relation for the secondary by \citet{ech83} we obtain a mass of $M_2 = 0.133\, \rm{M}_{\sun}$. Similar results are obtained using other calibrations: $M_2 = 0.128\, \rm{M}_{\sun}$ \citep{pat84}; and $M_2 = 0.152\, \rm{M}_{\sun}$ \citep{hnr01}. The secondary mass of 0.108 $\rm{M}_{\sun}$ found by \citet{beu08} indicates a lower mass than a main-sequence star for that orbital period. This is most probably the result of mass-loss of the companion star during the binary secular evolution. However, this mass, the spectral type found by these authors and the mass ratio obtained in this paper clearly indicates that the system is still approaching the bounce-back limit (see \citet{pat11} and references therein).

\subsection{Spin and orbital modulation of the broad component}
\label{sec:swidth}

\begin{figure}
 \begin{center}
\includegraphics[trim=0 0 0 0,clip,angle=270,width=\columnwidth]{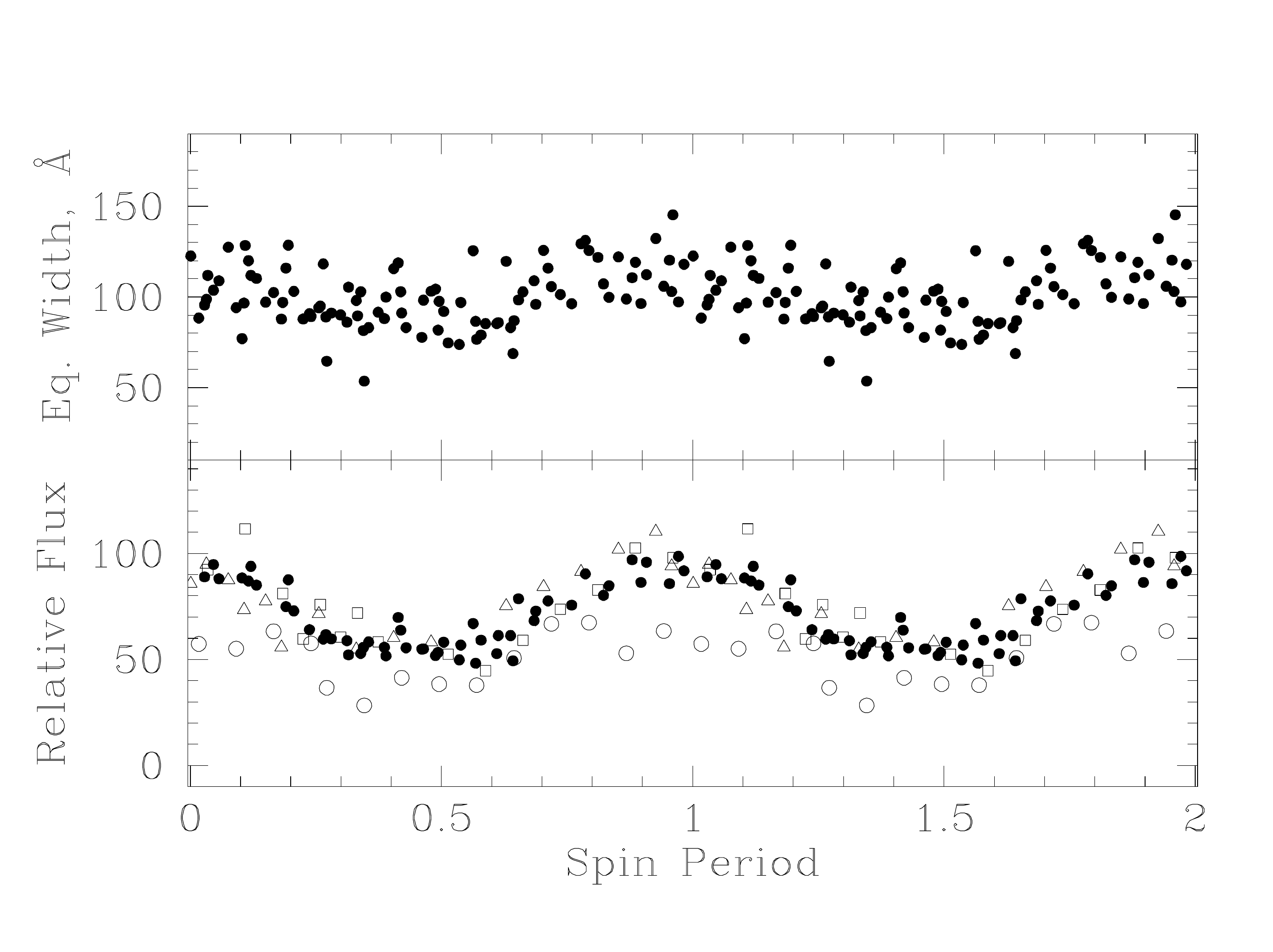} \\
\caption{Equivalent width (\emph{top})  and relative flux (\emph{bottom}) of the whole H$\alpha$ line folded on the 67-min spin cycle. There is a weak modulation of the equivalent width consistent with previous results. The relative flux of the line shows a stronger modulation with the spin period. However, the strength of this modulation depends not only on the spin activity but also on the corresponding orbital phases. Different open figures correspond to specific orbital phases (see text for details). }
\label{fig:mod_all}
\end{center}
\end{figure}

\citet{bre80,gil82,hel87} have shown that the Balmer lines show a modulation with the spin period. We have done a similar analysis for H$\alpha$ and found also a weak modulation of the equivalent width of the whole line with the spin cycle, as shown in the top panel in Figure~\ref{fig:mod_all}. There is an observed maxima at phase 1.0, whereas the minimum occurs at phase 0.5, consistent with previous results. The relative flux of the whole line (Figure~\ref{fig:mod_all}, \emph{bottom}) shows a stronger modulation with the spin period. However, the strength of this modulation depends not only on the spin activity but also on the corresponding orbital phases. To illustrate this, we have separated three spin cycles corresponding to: open triangles, first half and open squares, second half of the spectra, third night; open circles, first half of the night. Note, for example, that around spin phase zero, the triangles correspond to orbital phases close to 0.5 and have a high flux, while open circles correspond to orbital phases around 0.0 and have a lower flux. In this scenario, the magnetic pole has maximum exposure, but in the first case the secondary is behind the white dwarf, while in the second case it passes in front. 

\begin{figure}
 \begin{center}
\includegraphics[trim=0 0 0 -140,clip,angle=270,width=1.2\columnwidth]{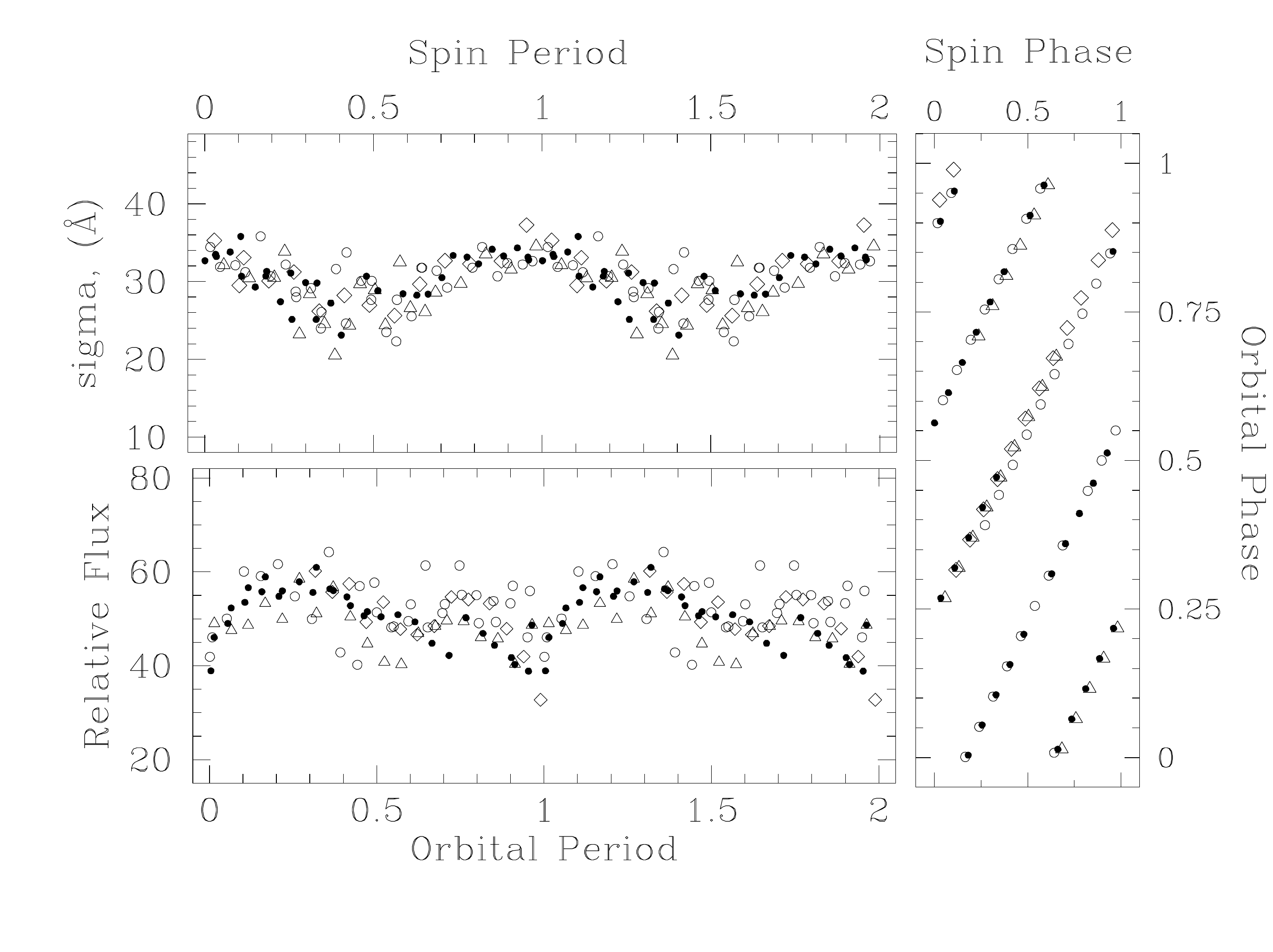} \\
\caption{The physical width of the line, or Gaussian sigma for the broad component of H$\alpha$ (\emph{top}) folded on the 67-min spin cycle. There is a clearer modulation on the wings of the emission line once the low-velocity regions have been masked. The calculated relative flux (\emph{bottom}) for the Gaussians fitting the broad component shows a small modulation and a sudden decrease at phase 1.0 when folded on the 98-min orbital cycle.  The relation between the spin and the orbital phases for the four observed nights are shown in the right panel.  The symbols in all panels are: open triangles, 1st night; open squares, 2nd night; solid circles. 3rd night and open circles, 4th night (see text for further details).}
\label{fig:mwings}
\end{center}
\end{figure}

Since we have also analysed the broad wings of H$\alpha$ alone, we have measured their Gaussian sigma. The results are shown in Figure~\ref{fig:mwings} (top panel). The spin modulation is now much clear once the low-velocity regions have been masked. Furthermore, we have calculated the relative flux of the fitted Gaussians (bottom panel) and found that this flux is modulated with the orbital period. There seems to be a weak modulation with maxima at orbital phases 0.25 and 0.75. In the right panel we have plotted the relations between the spin and orbital period phases, which are, in the case of EX~Hya tightly locked, because the close 2:3 relation. The symbols, which are the same as in the left top and bottom panels are described in the Figure caption. Further discussions on these modulations can be found in Section~\ref{sec:discus}.
 
\section{Doppler Tomography}
\label{sec:dopmap}

\begin{figure}
  \begin{center}
     \includegraphics[trim=0cm 0cm 0cm 0cm,clip,width=8cm]{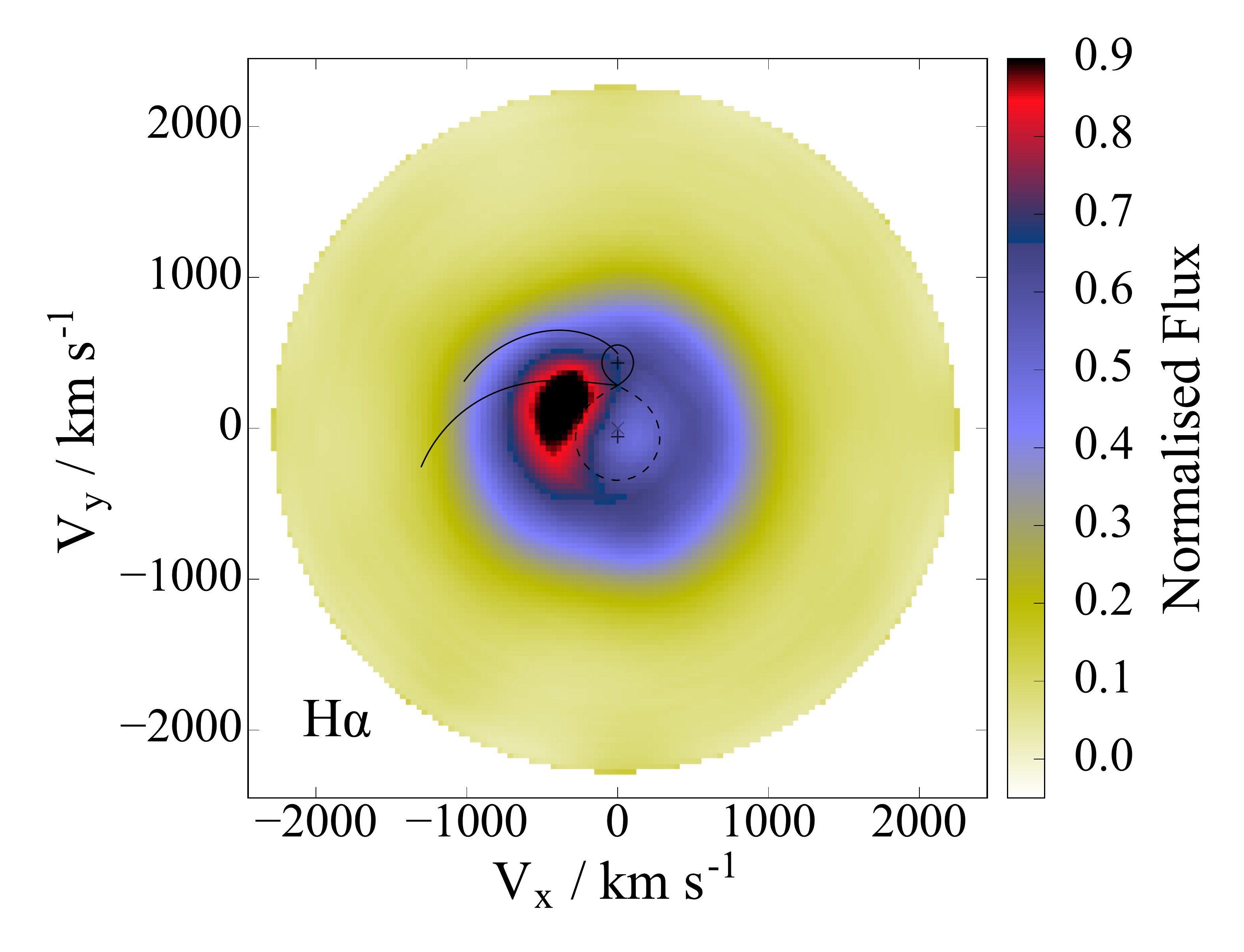}  \\
     \includegraphics[trim=0cm 0cm 0cm .4cm,clip,width=8cm]{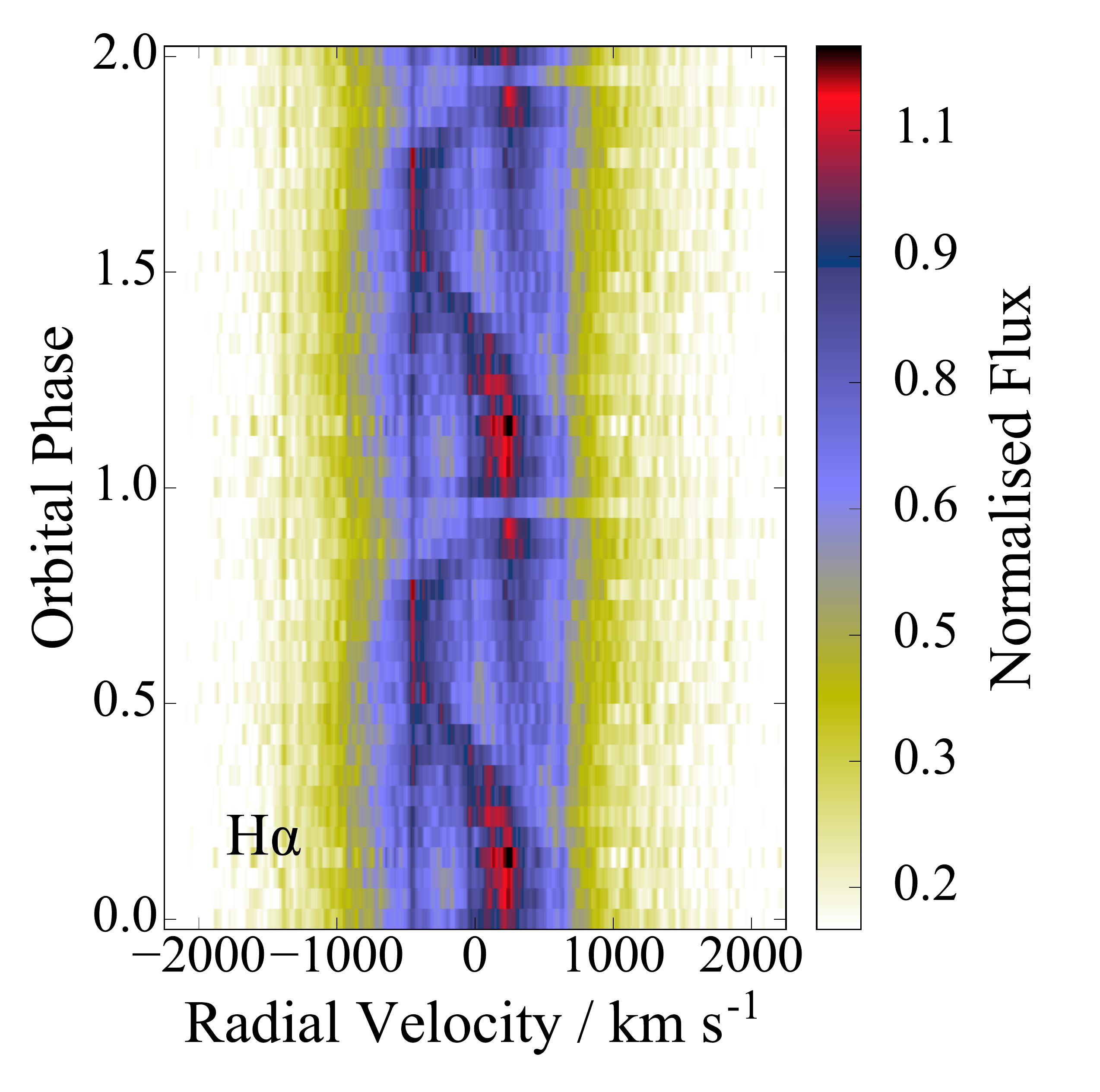}  
    \caption{Overall H$\alpha$ tomogram and trail spectra showing a narrow S-wave emission produced by a hotspot, a double-peaked emission mapped into an accretion ring with velocities up to $\sim$1000\kms and a high-velocity component extending to $\sim$2000\kms; component which, while smeared out in the outer regions of the tomogram, is clearly present in the trail spectra,  (see text for details and Figure~\ref{fig:em-lines} also).  The colour scale at the right shows the normalized intensity. }
     \label{fig:streamAll}
\end{center}
\end{figure}

Doppler tomography is a powerful technique used to unveil the features of accreted material in cataclysmic variables \citep{mar88}. We have constructed an overall tomogram and trail spectra for the H$\alpha$ emission line in Figure~\ref{fig:streamAll} to see the contribution of the three regions identified in Section~\ref{sec:comp}: a narrow S-wave emission; a double-peaked emission (presumably from the disc); and a weak broad component, usually interpreted as originating in the accretion curtain. In fact we see a strong asymmetric feature at low velocities, which is consistent with an S-wave produced by a hotspot. We also find an accretion disc component of intense H$\alpha$ emission with velocities up to $\sim$1000\kms, arising from an outer region to the inner disc, 
in which the disc is presumably disrupted by the magnetic field of the white dwarf. The third component is in fact a weak high velocity region, faintly observed in the tomogram up to $\sim$1500\kms. As stated by \citet{kot15}, in a standard projection, Doppler tomography tends to concentrate and enhance lower velocity features while higher velocity features are more separated and smeared out. Figure~\ref{fig:streamAll}, left panel shows the Doppler tomogram spanning velocities up to 2700\kms. Nevertheless, the outer region corresponding to the high-velocity component has been diluted and it shows a faint outer disc barely exceeding $\sim$1500\kms, while the trail spectra on the right panel better trace the wings of the line profile extending to velocities up to $\sim$2000\kms. The colour scale at the right of  the figure shows the normalized intensity used for the construction of the trail spectra and its tomogram. The high velocity component is also visible in the line profile of the co-added spectrum, particularly for H$\alpha$ (Figure~\ref{fig:em-lines}, top) which shows a change of slope at about $\sim$1000\kms from the line center and intensity keeps going down to reach the continuum at about $\sim$2100\kms on the blue wing and $\sim$2300\kms on the red wing. The overall trail spectra also shows a decrease of intensity in the low velocity regime around phase zero as supported by the flux of the broad component (Figure~\ref{fig:mwings}, bottom).

Since there are clear and strong photometric modulations with the spin period (see Figure~\ref{fig:vphot}), we have compared the original spectra tomograms with those obtained from normalized spectra, corrected with the simultaneous photometry. We found that the latter has little effect on the H$\alpha$ tomograms and we have therefore used the original spectra. This is not surprising as we have already discussed that high velocity components tend to be smeared out in the standard Doppler projection.

\begin{figure*}
  \begin{center}
  	\begin{tabular}{ccc}

    \includegraphics[trim=1cm 0cm 1.6cm 1cm,clip,height=4.5cm]{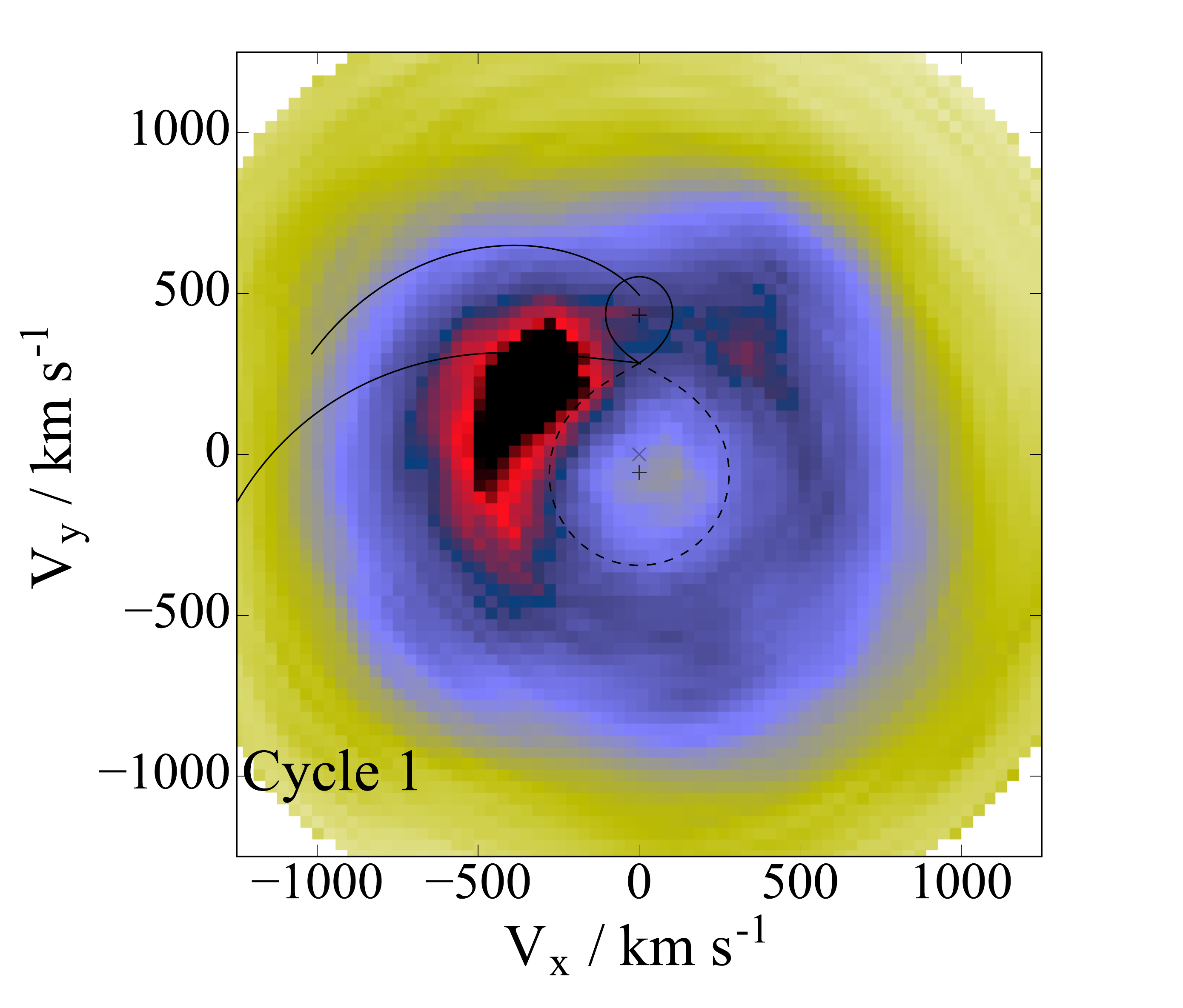}  &     
    \includegraphics[trim=1cm 0cm 1.6cm 1cm,clip,height=4.5cm]{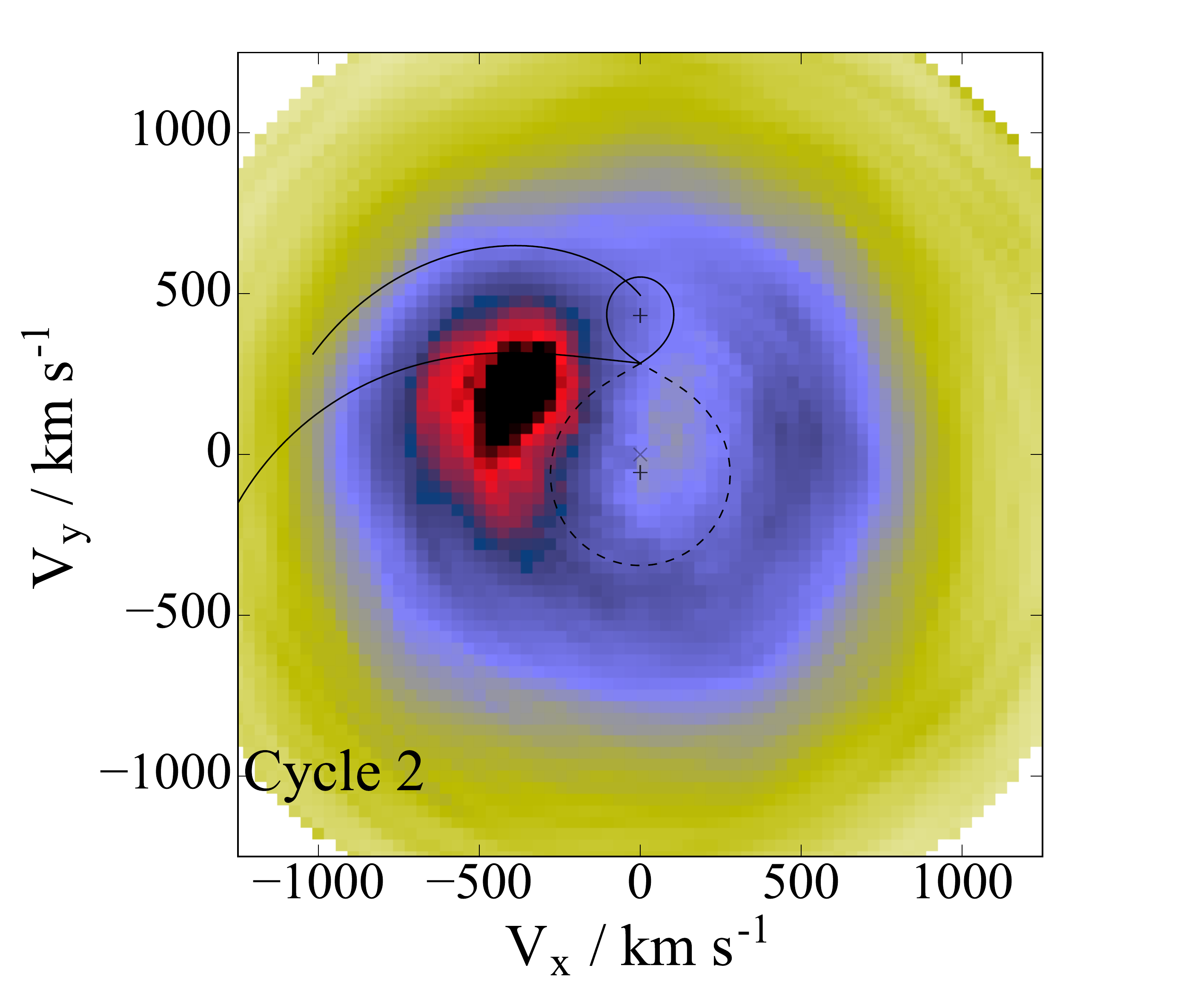}  &
    \includegraphics[trim=0cm 0cm 0.cm 0.8cm,clip,height=4.5cm]{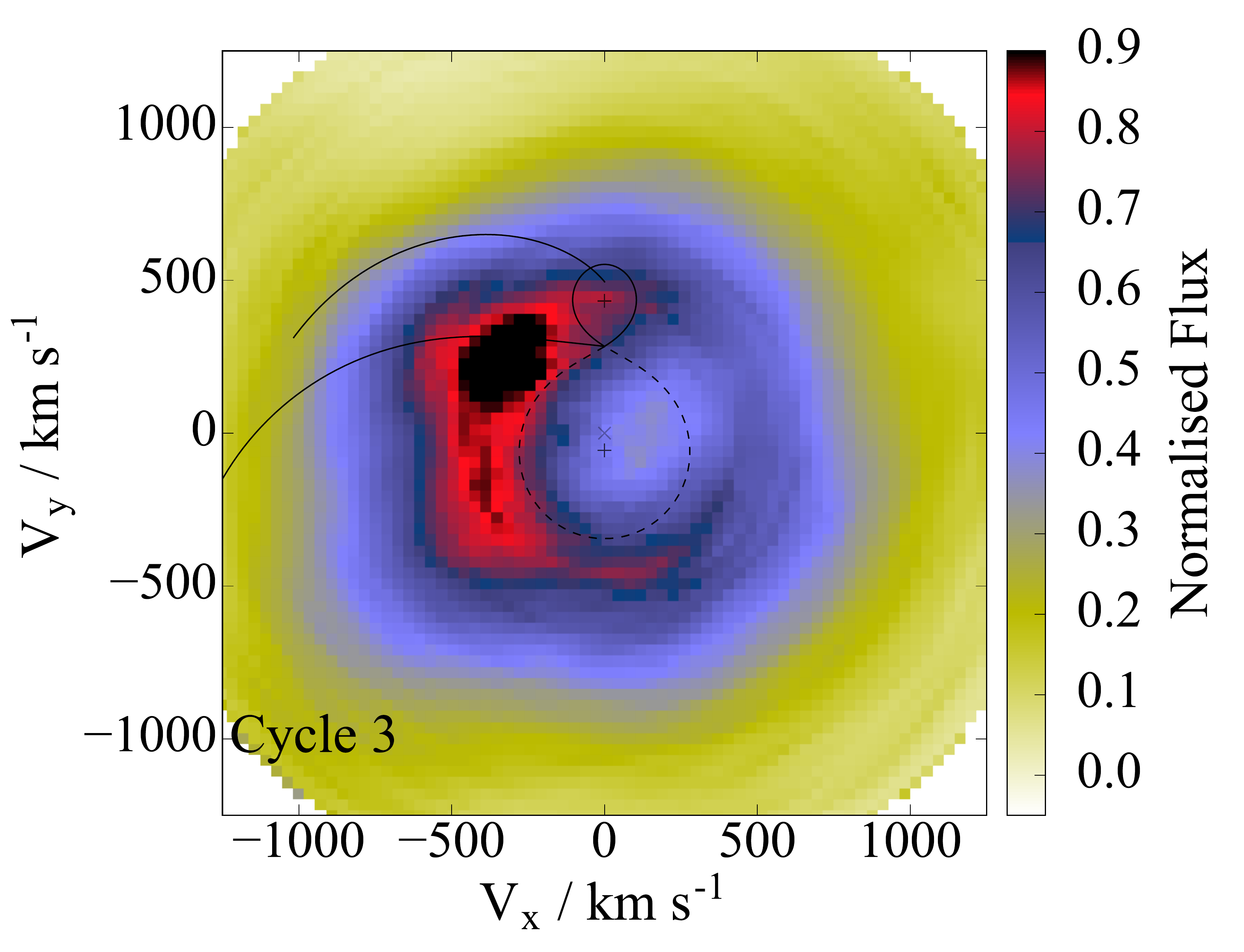} \\
    \includegraphics[trim=1cm 1.1cm 1.6cm 1cm,clip,height=4.5cm]{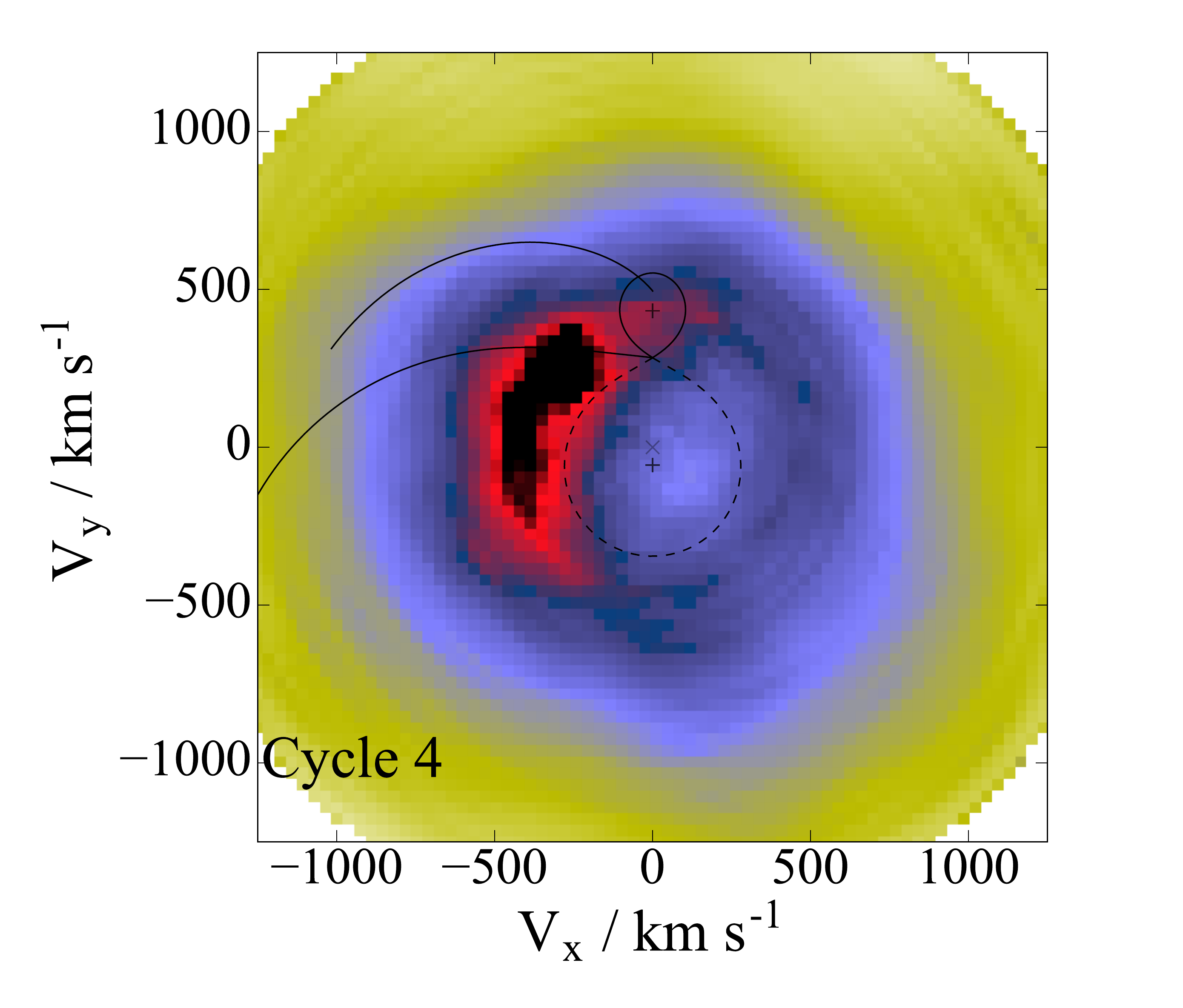}     & 
    \includegraphics[trim=1cm 1cm 1.6cm 1cm,clip,height=4.5cm]{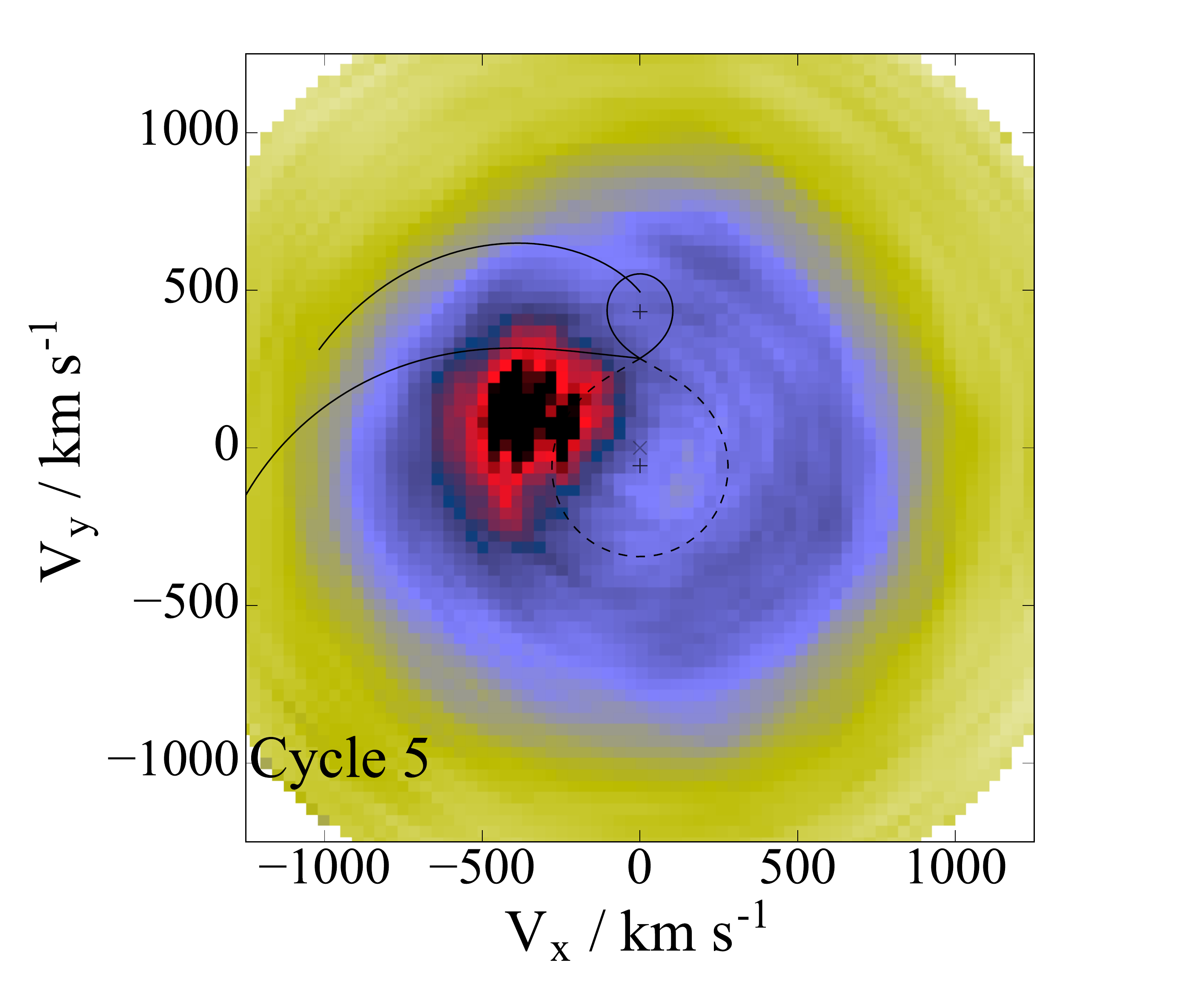}  &
    \includegraphics[trim=1cm 1cm 1.cm 0.8cm,clip,height=4.5cm]{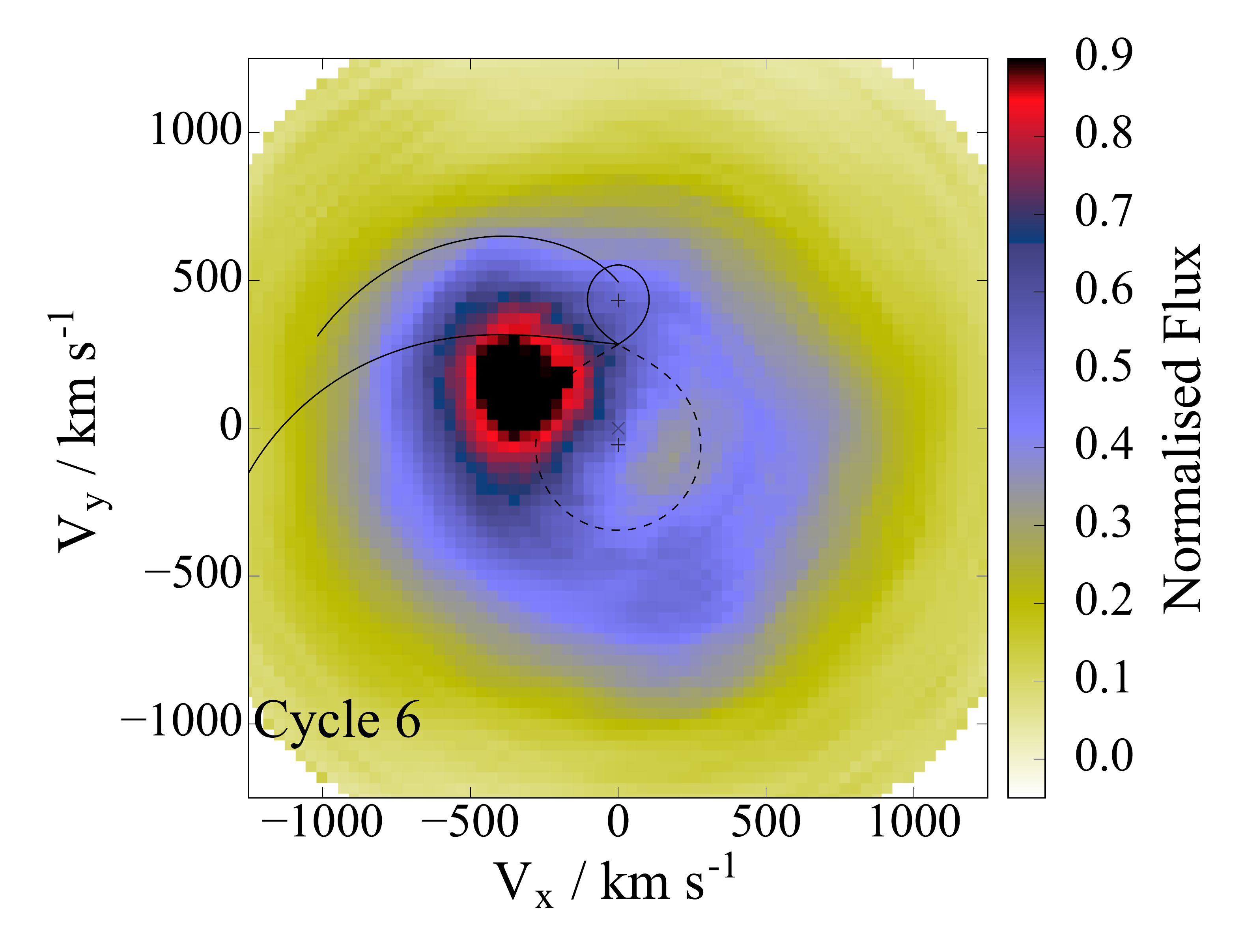}  \\ 
    \end{tabular}
    \caption{H$\alpha$ Doppler tomograms for each orbital cycle (1-6, from top left to bottom right). We have restricted the analysis to a region of $\leqslant$1600\kms. The projected velocity of the center of mass of each component (\textit{crosses}) and the common center of mass (\textit{plus}), as well as the Roche-lobe surface of the secondary are shown for reference. The Keplerian (\textit{top line}) and ballistic trajectories (\textit{bottom line}) are also shown. We observe a strong variable hotspot, which sometimes has stream-like feature, associated with ejection and accretion episodes. We also see a well formed disc component (see text for more details).}
    \label{fig:tomophot}
  \end{center}
\end{figure*}

\begin{figure*}
  \begin{center}
    \includegraphics[trim=0 0 0 0,clip,width=.60\columnwidth]{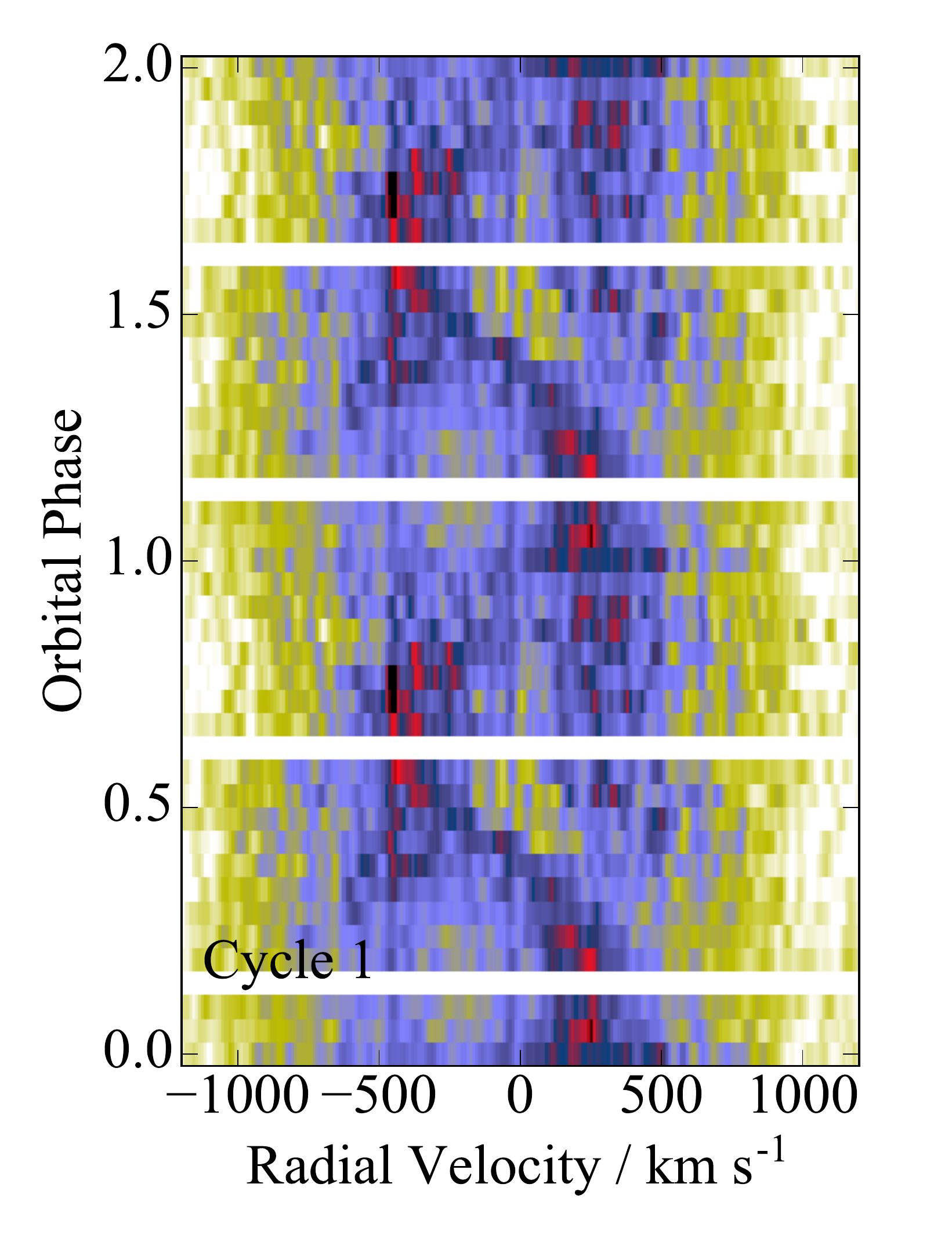}       
    \includegraphics[trim=0 0 0 0,clip,width=.60\columnwidth]{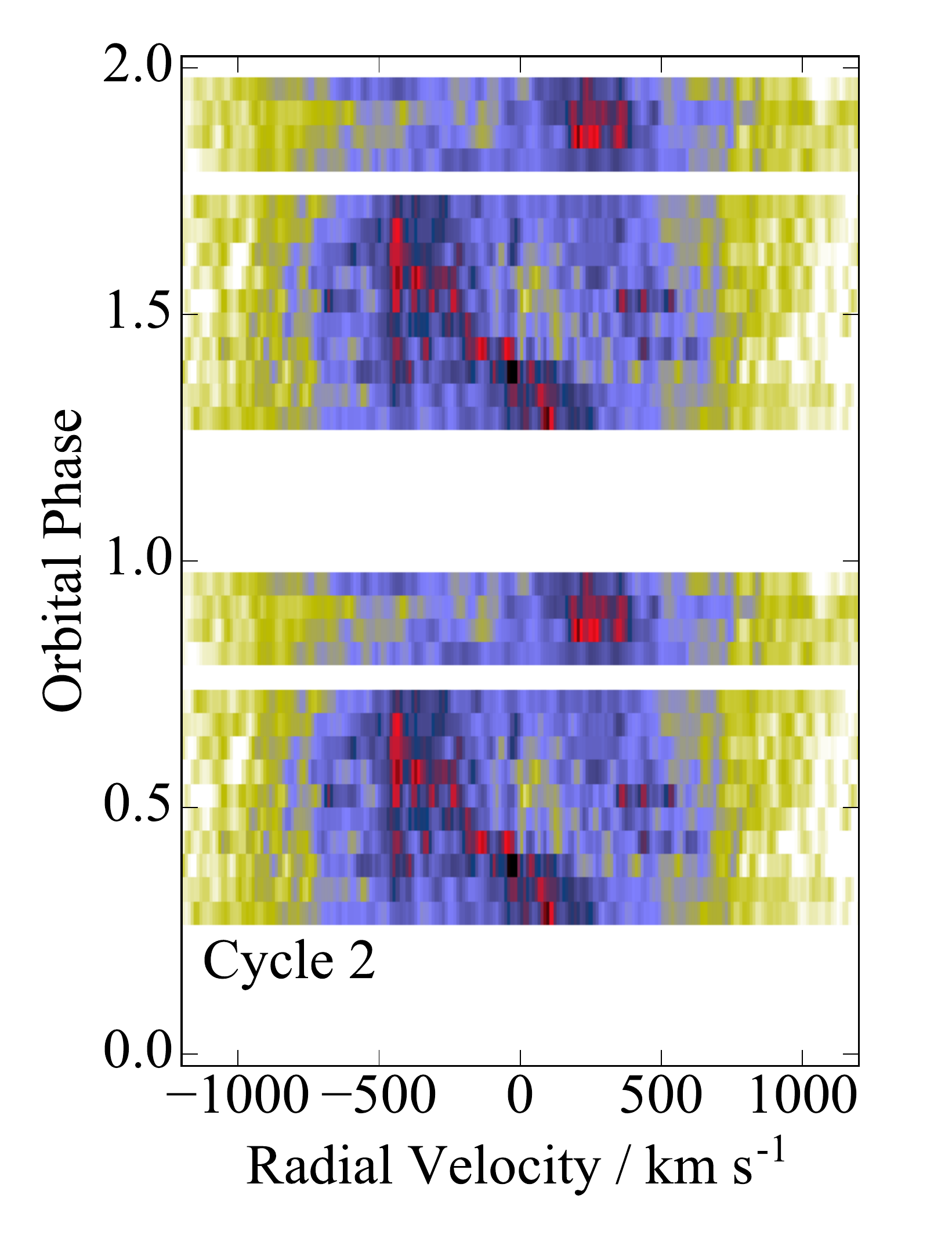}  
    \includegraphics[trim=0 0 0 0,clip,height=0.78\columnwidth]{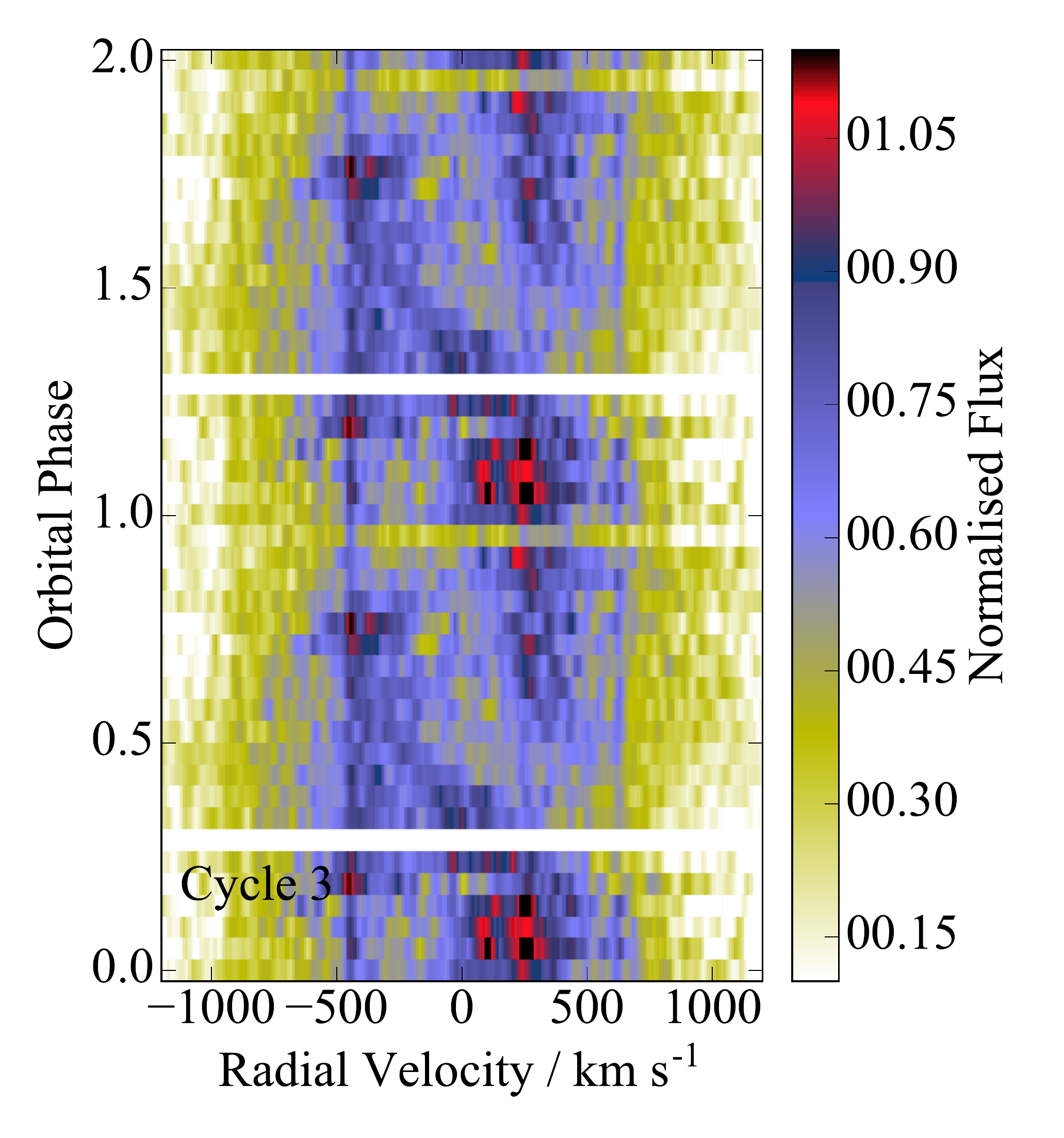} \\
    \includegraphics[trim=0 0 0 0,clip,width=.60\columnwidth]{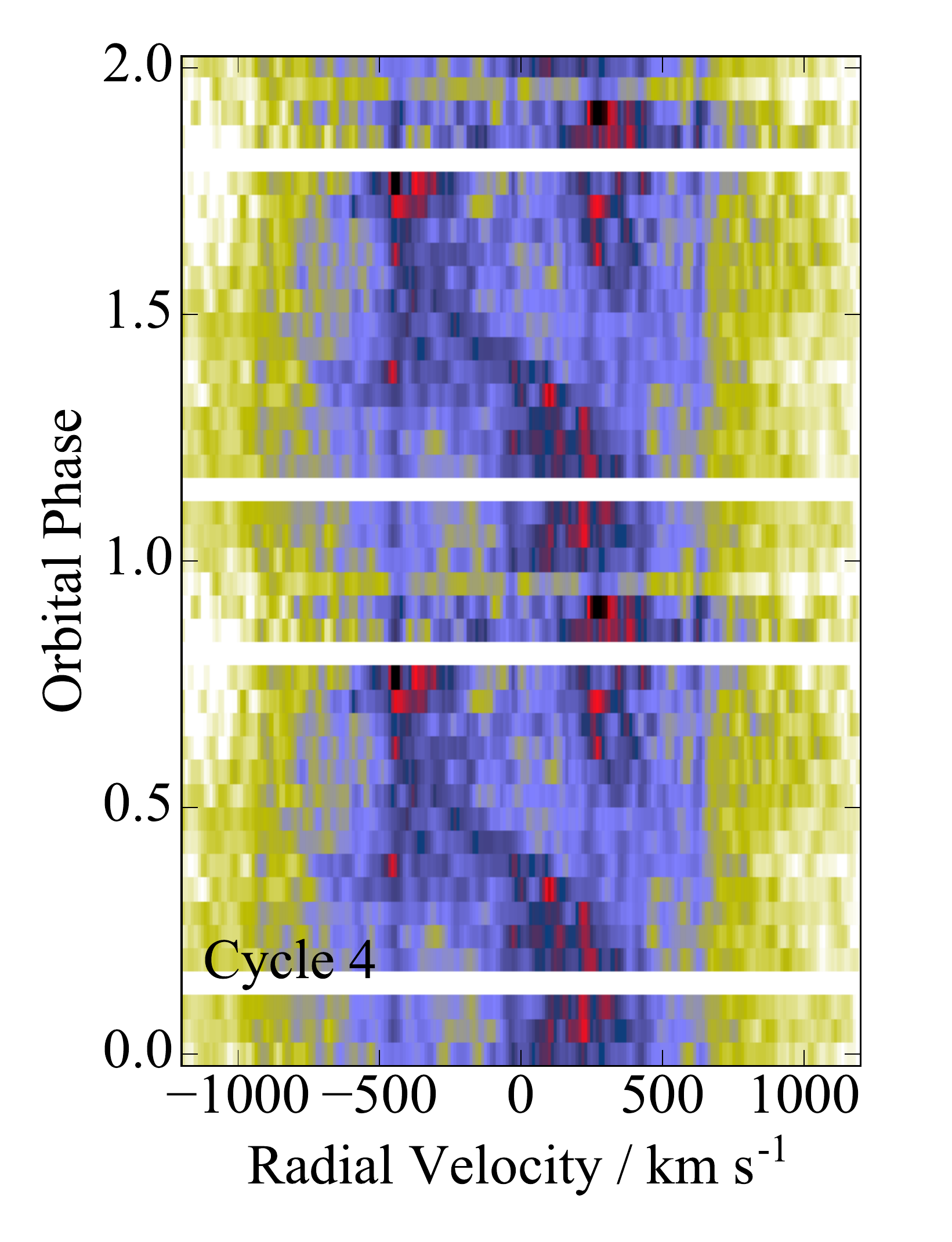}     
    \includegraphics[trim=0 0 0 0,clip,width=.60\columnwidth]{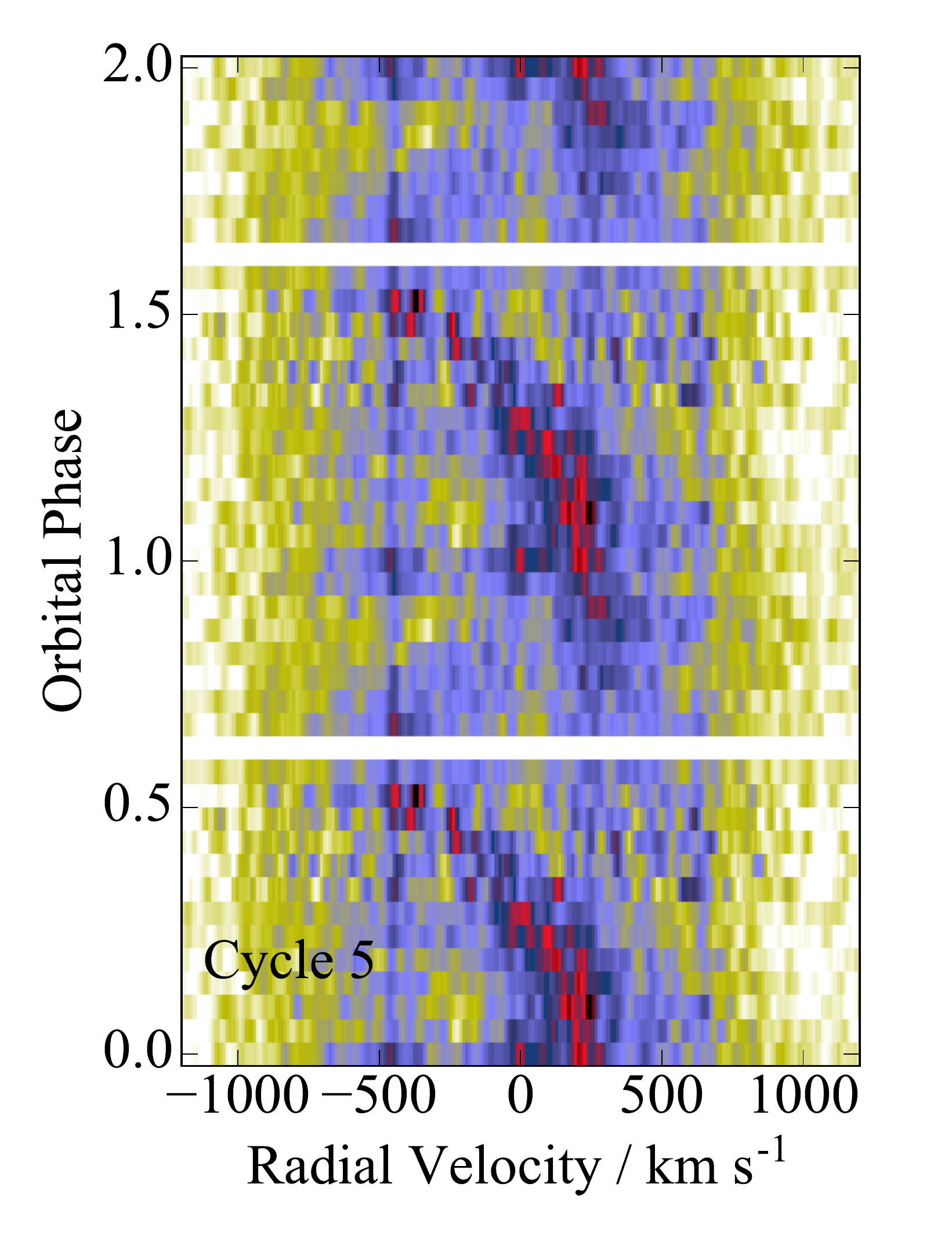} 
    \includegraphics[trim=0 0 0 0,clip,width=.73\columnwidth]{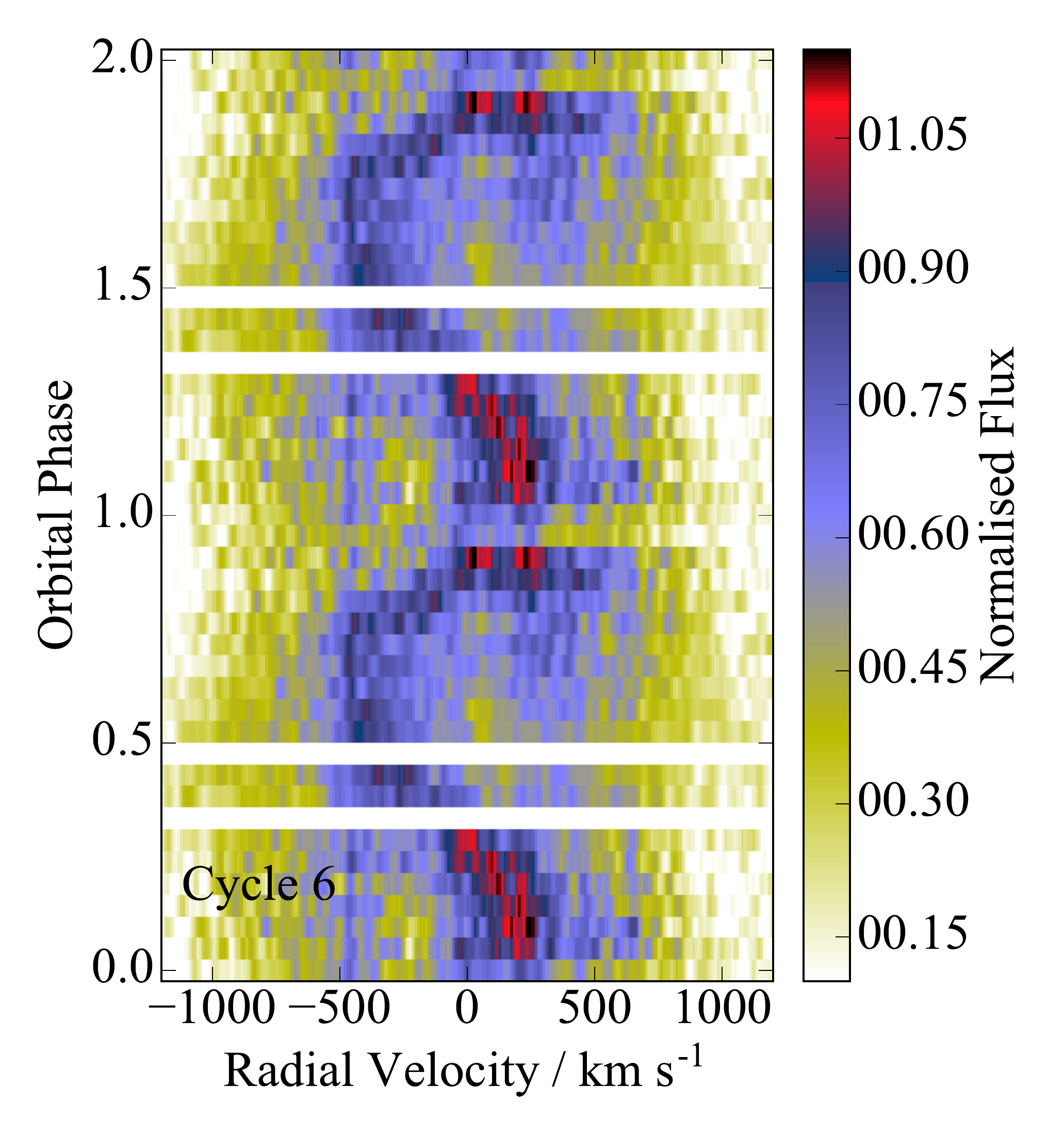}  \\ 
    \caption{H$\alpha$ trail spectra as a function of orbital phase corresponding to the orbital tomograms 
    in Fig~\ref{fig:tomophot}. The horizontal axis is the projected velocity in \kms. There seems to be an eclipse of the S-wave near phase zero in most of the trail spectra. The double peak, the S-wave and other features are discussed in the text. }
    \label{fig:trailspec}
  \end{center}
\end{figure*}

As there is no substantial information of the high velocity region in our tomograms, in contrast with the radial velocity results presented in Section~\ref{sec:radvel2}, we decided to focus our analysis on the velocity region  $\leqslant$1300\kms.  Since we have enough information to construct single orbital cycle tomograms during  our observational run, we present six H$\alpha$ tomograms, in order to detect variations on the hotspot and the accretion  region up to 1300\kms. During the first two nights we have  covered only a single orbit, while for the last two nights we have completed about 1.5 orbital periods, which we have divided into two tomograms overlapping in time, as indicated by the bars depicted under in the simultaneous light curves shown in Figure~\ref{fig:vphot}. Hereinafter referred as tomograms 1 to 6 in Figure~\ref{fig:tomophot}: from top left, the first night; bottom left, second night; middle panels, third night and right panels, fourth night. The corresponding trail and reconstructed spectra are shown in Figure~\ref{fig:trailspec} following the same order.

Our results from the individual Doppler tomograms are similar to the overall tomogram shown in Figure~\ref{fig:streamAll}. However, we detect substantial variations, particularly on the shape of the strong hotspot. In most nights it appears well constrained and near the $L_1$ point, while in other cycles it appears highly smeared. This smeared feature, or stream, could be correlated to a different combination of spin-orbital cycles. Further discussion can be found in Section~\ref{sec:discus}.

The trail and reconstructed spectra in Figure~\ref{fig:trailspec} are dominated by the strong hotspot and the stream. The double-peaked accretion disc appears sometimes in the background with a peak-to-peak separation of $\sim$1000 km~s$^{-1}$. The contribution from the hotspot alone is clear in the narrow S-wave in the first and fourth nights and it shows a variation with an amplitude of $\sim$500  km~s$^{-1}$. There seems to be an eclipse of the S-wave near phase zero in most of the trail spectra, nevertheless in trail 3 and 4 the eclipse appears to affect the red-shifted wing, while in trail 2 and 5 there is almost no eclipse at all. Furthermore, there is a bright spot in trail 2 near the S-wave. The stream is clearly shown in nights two and three, where we see a broader S-wave. This stream is particularly strong, both in tomogram 5 and also in its corresponding trail spectra, where there is also a prominent contribution at phases 0.1 to 0.3. There is a red-shifted emission at phase 0.6 in trail 1 to 4, extending from a slightly greater velocity amplitude than the hotspot. This emission does not follow the double-peaked Doppler shift and it encounters the red wing of the hotspot at phase 1.0. A blue-shifted counterpart, occurring shorter in phase, is observed around phase 0.3 in trail 1 and 5, and around phase 0.4 in trail 4. These features, combined with our radial velocity results, will be discussed in the next section.

\section{Discussion}
\label{sec:discus}

\begin{figure}
 \begin{center}
\includegraphics[trim=0cm 10cm 0cm 6cm,clip,width=1.0\columnwidth]{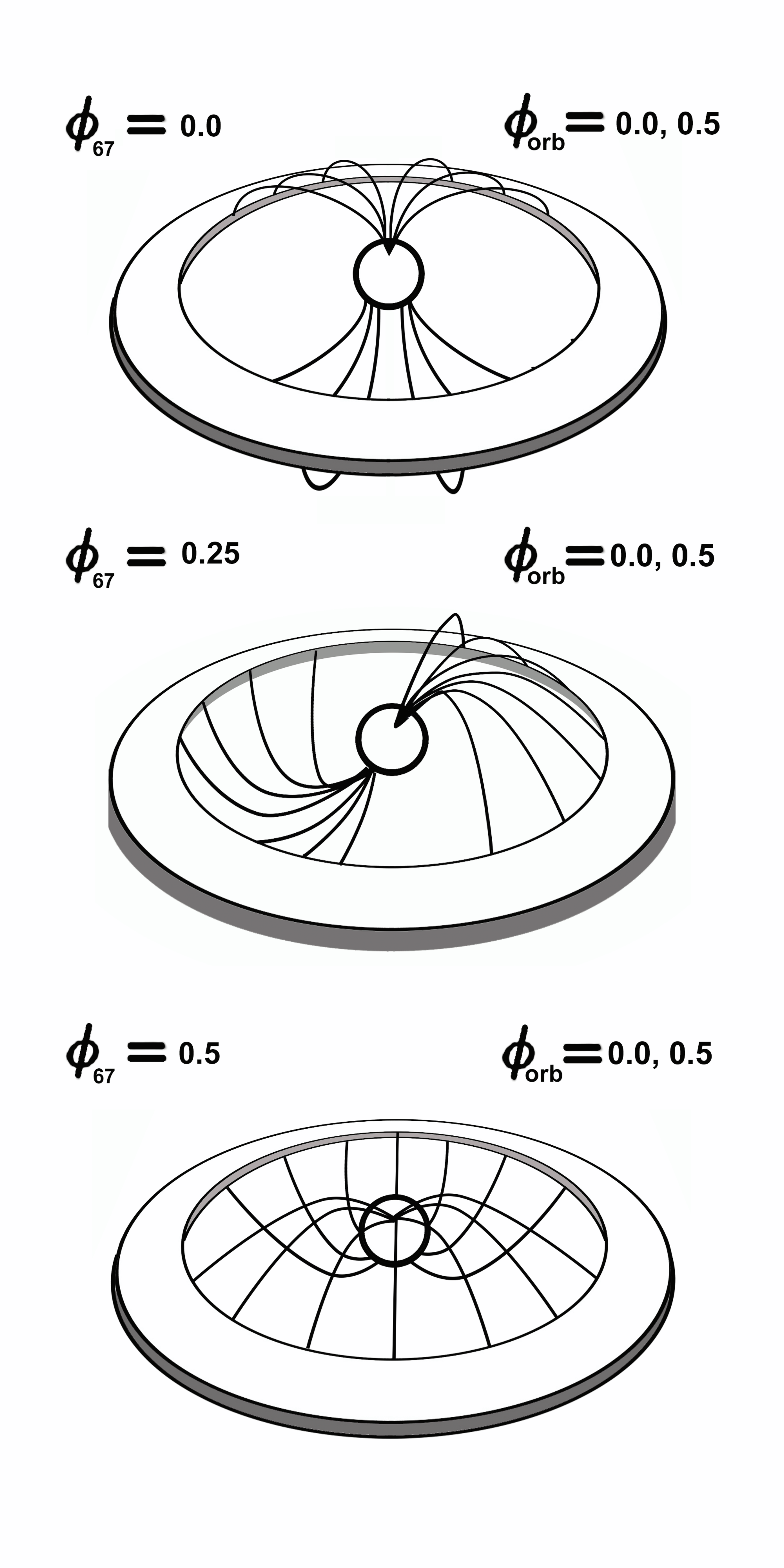}
\caption{Diagram of the inner disc and the magnetosphere at different phases. The observer is in front for orbital phase 0.0 and behind at phase 0.5 in all cases (see text).}
\label{fig:diagram}
\end{center}
\end{figure}

EX~Hya is a well studied system, both photometrically and spectroscopically. However, its spin to orbital period ratio, close to 2:3, complicates its analysis and creates controversies, leaving still some unsolved problems. Among them is the existence or not of an accretion disc, and consequently the question of the size of the inner disc and the magnitude of its magnetosphere \citep{hel14,mhl07a}. Another remaining problem is the constancy of the orbital period. While we have shown here that the orbital period remains constant, we can not rule out a possible small sinusoidal oscillation (as far as our own observations). However the spin period is definitely decreasing at a slow rate \citep{mau09}. Another important problem is the accurate determination of $K_1$ and the mass of the WD (even in this eclipsing system), as we discussed in Section~\ref{sec:spec}. 

With respect to the problem of the radial velocity semi-amplitudes of the accretion disc (and presumably of the primary star), we have pointed out that the published optical data have a large scatter (see Table~\ref{tab:k1lit}). These works have been performed in the blue part of the visible spectra, while our observations are centred around a strong H$\alpha$ emission line. It is plausible that the disc component is more dominant in H$\alpha$ than in higher order Balmer lines and hence, the $K_1$ result would give a more accurate determination. The similarity of the results in the radial velocity curves (see Section~\ref{sec:spec}), obtained from medium and high velocity regions have a deep implication: the high velocity component that we have analysed is moving in the orbital plane and therefore should be interpreted as an inner disc region. In fact, our analysis indicate a robust result, both from the main disc component as well as from the broad wings. Furthermore, our $K_1 = 58 \pm 5$\kms value is in accord with the UV \citep{bel03} and X-ray \citep{hoo04} results. 

Still, the observed RM-type effect remains a problem and defies easy explanation. If this is the selective occultation of a rotating body, the simple explanation would be that the secondary star is partially occulting a symmetric disc. Before (but close to) phase zero, the secondary star will (partially) occult first the blue shifted parts of the disc, while shortly afterwards, it will mask the receding parts of the disc. As a result, the radial velocity distortions should be centred around phase zero, and this should be observed in both, the main disc and the inner disc. We observe, however, that the asymmetry in both cases, is centred around phase 0.9. This implies that there is an asymmetric component present, and there is only one asymmetric component in the disc that could account for this shift, namely, the hot spot. We have calculated that the phase offset $\phi_o$ between eclipses and inferior conjunction is around 0.1 (see Table~\ref{tab:OrbParam2}). Although we have avoided measuring the hot spot velocity component, it is possible that the hot spot itself is physically occulting the high velocities behind it. Therefore, we propose that the RM-type effect is in this case a combination of the occultation of the disc velocity regions by both the hot spot and the secondary star. This explanation is supported by the position of the spot shown in the tomography and by the fact that the RM-type effect moves closer to phase zero for the broad wings (see Figure~\ref{fig:velcur2}), suggesting that the effect of the occultation  by the hot spot is smaller for higher velocity regions. In other words, the hot spot eclipse is a grazing eclipse and does not affect the central parts of the disc, nor the white dwarf itself.

With respect to the accurate determination of the masses of the binary, we believe that our careful determination of $K_1$ by two different methods, consistent with UV and X-ray measurements, coupled with the accurate determination of $K_2$ by \citet{beu08}, enable us to derive reliable masses (see Section~\ref{sec:baspar}). The obtained values are consistent with a short period system with a primary mass with a mean value of 0.8 M$_{\sun}$ (e.g. \citet{pat11}) and a secondary star with enough mass to produce nuclear reactions, still approaching the minimum orbital period \citep{kab99}. A long standing discrepancy with the mass of the primary obtained indirectly by X-rays, has also been resolved recently by \citet{lea15}.
 
The results of the Doppler tomography are quite revealing. Although the overall tomogram, shown in Fig~\ref{fig:streamAll}, does not reveal the weak broad component, due to the dilution effect in a standard projection, mentioned in Section~\ref{sec:dopmap}, the trail spectra does show this high velocity component (up to $\sim$2000 \kms). The individual tomograms, show that there are fast changes in the hot spot or stream from cycle to cycle, and particularly from night to night, revealing different combination of spin-orbital cycles. To illustrate the effects of the spin-orbital configuration in the tomography, as well as in other related aspects discussed below, we present in Figure~\ref{fig:diagram}, three diagrams of the inner disc and the magnetosphere in a classical configuration (e.g. \citep{hel14}). (We must clarify at this point that these schematics are a visual help only and we do not intend to quantify any physical parameter from it). Spin phases 0.0 and 0.5 are defined at the time when the visible magnetic pole points away and towards the observer respectively. Each one has also been labelled for orbital phases 0.0 (when the secondary star is in front of the observer) and 0.5 (when the secondary is behind the disc). The hot spot and stream episodes appear connected with the spin-orbital cycle combination. Two examples are selected, which have the desired photometric combination of spin and orbital phases (see Figure~\ref{fig:vphot}). An episode with a more pronounced stream would correspond, as seen mainly in tomogram 4, when the material is smeared because the magnetic pole points toward the secondary star (\emph{lower diagram}, $\phi_{spin} = 0.5$, $\phi_{orb} = 0.0$); a less pronounced stream, e.g. tomogram 1, would produce a more compact spot, when the pole points away from the observer  but the secondary star is again at inferior conjunction (\emph{top diagram}, $\phi_{spin} = 0.0$, $\phi_{orb} = 0.0$). All other tomograms show, either a contained hot spot or a hot spot and a stream, but in these cases, although the spin and orbital phases are at some point in conjunction or opposition, the photometry has a complex behaviour, which makes the interpretation of these streams or compact spots more difficult. This would be the case of tomogram 3, a stream seems to dominate due to the high photometric activity near spin cycle 42. It is clear that these stream or compact spots do not affect our radial velocity results, as the hot spot and or stream, have been avoided by our double Gaussian of the main disc component and single Gaussian analysis of the broad component.

Although the H$\alpha$ radial velocity curves depicted in Figures~\ref{fig:velcur1} and \ref{fig:velcur2} have a clear orbital modulation, the spin and orbital modulations have a mixed behaviour. For example, the EW and more clearly, the relative flux, is modulated mainly with the spin cycle, but this modulation also depends on the orbital phase. As an example, we show in Figure~\ref{fig:mod_all} (lower panel) that the maximum flux occurs when the magnetic upper pole of the WD points away from the observer, at spin phase 1.0 and the secondary is behind the disc, while a minimum is seen at the same phase when the secondary is in front (Figure~\ref{fig:mwings}, top panel).

Our analysis of the broad component presented in Sections \ref{sec:radvel2} and \ref{sec:swidth} shows that the radial velocity behaviour of the inner region is very similar to that of the disc component. This suggests that the observed broad wings arise from material predominantly coming from the inner disc region, still in the orbital plane, although there may be also some small contributions from the accretion curtain. As we described earlier, the width of the broad component is modulated with the spin period (Figure~\ref{fig:mwings}, top panel). We interpret the maximum broadening around phase zero as a preferential fan-like view of the inner accretion curtain, when the magnetic upper pole points away from the observer. The grazing eclipse of the secondary star would not affect this broadening (see Figure~\ref{fig:diagram}, top diagram) as it will cover the lower magnetic pole only. However, the behavior of the relative flux of the broad line is not like the one we measure for the whole line (see Figure~\ref{fig:mod_all}, lower panel). In Section~\ref{sec:swidth} we explained that, although the flux is modulated by the spin-cycle, there is a clear correlation when the spin and orbital phases coincide. In the case of the broad wings, there is no correlation with the spin cycle (not shown in the paper), but there is an apparent double modulation with the orbital period. We must point out again, that the individual nights show different results, again depending on the correlation between the spin and the orbital phases, shown in the right panel of Figure~\ref{fig:mwings}. The double modulation is clearly seen in the first night, where the spin and the orbital phases do not coincide at phase zero. However, in night two, the flux is dominated by an eclipse at phase zero, when both spin and orbital phases coincide. Although there seems to be a maximum at phase 0.75, we do not have sufficient orbital coverage on that night to claim the double modulation. Where as in night three, which covers two spin cycles, there is a single modulation with a maximum at phase 0.25 and a shallow eclipse at phase zero. In night four, which also covers two spin cycles, there is only a large scatter in flux values. We also observe that the relative flux of the broad component is more sensitive to this spin-orbital correlation, than the flux observed for the whole line. We are uncertain, therefore, that the double modulation in flux is a real and stable feature. We believe that more observations are needed to confirm it. 

Our radial velocity study agrees with the main results by \citet{hel87}, yielding similar masses for the binary. But their range in $K_1$ values implies a primary mass between 0.62 and 0.99 M$_{\sun}$ with a most likely value of 0.78 M$_{\sun}$. The mass of the secondary star is obtained indirectly using an empirical ZAMS calibration by \citet{pat84}. The fact that the masses are in agreement is due to the use of our smaller $K_1$ and to the adopted value for $K_2$ from \citet{beu08}. Nevertheless, the mass result of the primary in both cases is close to 0.8 M$_{\sun}$, and it is also encouraging that the new cooling flow models from the X-ray emission, now agrees with this mass \citep{lea15}.

The eclipses seen in the trail spectra favour the interpretation of the S-wave component as a bright emission region near the outer edge of the circulating disc as stated by \citet{hel87}. The S-wave asymmetry, noted by these authors, is seen, but only in our trail spectra 1 and 3 with a more rapid transition from blue to red. The rest show a more sinusoidal behaviour as seen in trail 4, 5 and 6. Trail 2 has a gap between phases 0.1 and 0.3 making difficult to tell. We should note that our orbital phases are 0.1 ahead of those of \citet{hel92} due to the use of our new ephemeris. This creates only a shift and have no influence on the asymmetries or lack of them. The tail-like features below the S-wave, seen in most of our trail spectra extending to slightly higher velocities, are also observed by \citet{mhl07a}. They interpret these as a high velocity component caused by an overflow stream,  reaching $\sim 1000$\kms, although, in our case, they only reach velocities up to 600\kms. In addition, our reconstructed spectra 5, suggest that these tails may form a second S-wave, shifted 0.5 in phase with the main S-wave, in contrast with a 0.2 shift observed in the reconstructed spectra by \citet{mhl07a}. The reconstructed spectra  3 and 4 show a broader S-wave, in accordance with this stream, but still with velocities no greater than 600\kms. 

\citet{hel14} estimates a magnetospheric radius of about 4 white-dwarf radii  ($R_{\rm{WD}}$) from the eclipse timing analysis which locates the centroid of the spin-varying emission \citep{sie89}. We have to point out that the spin-varying emission does not hold in the long term (Figure~\ref{fig:omc}, \emph{bottom}). We believe that this effect and the possible sinusoidal variations with the orbital period might be due to a changing geometry of the system concerning the material locked by the magnetosphere and to the position of the hot spot. Nevertheless, our H$\alpha$ emission is strong in a region up to $\sim$1000\kms, and this would result in a minimum accretion disc radius of $R_{\rm{in}}\sim15\,R_{\rm{WD}}$, while the circularisation radius for a system with $q = 0.13$ \citep[Eq. 4]{kin99} turns out to be $R_{\rm{circ}}\sim18\,R_{\rm{WD}}$. However, if our high velocity component comes from a region with Keplerian velocity (i.e. an inner disc component still in the orbital plane), then the minimum disc radius would extend only to $3.75\,R_{\rm{WD}}$. This is in accordance with the fact that EX~Hya shows no polarisation, which is unlikely if the magnetosphere is to dominate the white-dwarf's Roche Lobe \citep{hel14}. Thus, our results point EX~Hya to have a small magnetosphere and being far from equilibrium. 

Although we have here claimed that the broad component comes from a disc-like structure, why does the emission line profile have an inflection point around 1000 \kms? And why is this component modulated mainly by the orbital modulation and not by the spin modulation as in the case of the whole line? Does this inflection point affects our result on our calculation of the inner radius of the disk? These are questions that need to be raised for future studies of EX~Hya; further observations are suggested to look closely at these points.

\section{Conclusions}
\label{sec:conclusions}

\begin{enumerate}

\item Our photometric observations of EX~Hya show a strong modulation with the 67 min spin period and 0.4\,mag narrow eclipses associated with the 98 min orbital period.  

\item We have computed new ephemeris for EX~Hya:  $$HJD_{\rm{eclipse}} = 2,437,699.94131(11) + 0.068233843(1) E.$$

\item The radial velocity analysis yields a semi-amplitude $K_1= 58 \pm8$~km~s$^{-1}$, both from the disc and broad components which, combined with the $K_2$ value from \citet{beu08} and an inclination angle of $i =78\degr \pm 1$, agreed by most authors, gives a determination of  $M_{1} = 0.78 \pm 0.03$ M$_{\sun}$, $ M_{2} = 0.10 \pm 0.02$ M$_{\sun}$ and $a = 0.67 \pm 0.01$ R$_{\sun}$.  

\item The equivalent width of the whole line is slightly modulated with the spin period as found by previous authors. Furthermore, we find that its relative flux is strongly modulated with the spin period, but their maxima and minima, depend on the alignment of the spin pole and the inferior conjunction of the secondary star.

\item The Gaussian sigma of the fit of the broad component is also clearly modulated with the spin period, even more clearly than the equivalent width of the whole emission line. There is a maximum at phase 1 and a minimum at phase 0.5. We interpret this as a fan-like view of the inner accretion curtain, when the magnetic upper pole points away from the observer. However, the relative flux of this broad component shows an apparent double modulation with the orbital period. Due to the dissimilarities in the individual nights, we are uncertain that this modulation is a real and stable feature and believe that more observations are needed to confirm it. 

\item We propose that the RM-type effect is the result of the combination of the occultation of the disc velocity regions by both the hot spot and the secondary star; an explanation which could account for the varying phase shift of the distortion.

\item The Doppler tomography reveals a well formed disc and a strong hot-spot, at times smeared. The changes in shape of the hot-spot could be due to a different combination of spin-orbital cycles.

\item The observational analysis we have made in this paper, including the radial velocity study, the simultaneous photometry and the Doppler tomography, indicate that EX~Hya has a well formed disc and an inner disc region, both which shows the same radial velocity behaviour. There is also a an indication of a small contribution from the accretion curtain. From the velocity of the broad wings, we estimate that the magnetosphere should extend only to about $3.75\,R_{\rm{WD}}$.
 
 \end{enumerate}

\section*{Acknowledgements}

The authors are indebted from DGAPA (Universidad Nacional Aut\'onoma de M\'exico) support, PAPIIT projects IN111713 and IN122409. This article is based upon observations collected at the Observatorio Astron\'omico Nacional at San Pedro M\'artir, B.C., M\'exico. We also thank the staff of the Observatory for their invaluable help and to Juan Carlos Yustis for producing the inner disc diagram.  JVHS acknowledges support via studentships from CONACyT (M\'exico) and the University of Southampton. We would like to thank the anonymous referee for the prompt response and excellent feedback and to Rafael Costero for his invaluable insight into the physics of this system.



\bsp	
\label{lastpage}
\end{document}